\documentclass[ reprint,
superscriptaddress,
 amsmath,amssymb,
 aps,
prb,
]{revtex4-1}
\usepackage{graphicx}
\usepackage{amsmath}
\usepackage{placeins}
\usepackage{color}
\usepackage{bm}
\newcommand{\SSS}{\mathbf{S}}

\newcommand{\BBB}{\mathbf{B}}

\newcommand{\zzz}{\mathbf{z}}
\newcommand{\mmm}{\mathbf{m}}

\newcommand{\Lutts}{Lu$_2$Ir$_2$O$_7${}}
\newcommand{\Ndtts}{Nd$_2$Ir$_2$O$_7${}}

\newcommand{\Ybtts}{Yb$_2$Ir$_2$O$_7${}}
\newcommand{\Tbtts}{Tb$_2$Ir$_2$O$_7${}}
\newcommand{\Gdtts}{Gd$_2$Ir$_2$O$_7${}}
\newcommand{\Hotts}{Ho$_2$Ir$_2$O$_7${}}

\newcommand{\YbTitts}{Yb$_2$Ti$_2$O$_7${}}

\newcommand{\kkk}{\mathbf{k}}

\newcount\colveccount
\newcommand*\colvec[1]{
        \global\colveccount#1
        \begin{pmatrix}
        \colvecnext
}
\def\colvecnext#1{
        #1
        \global\advance\colveccount-1
        \ifnum\colveccount>0
                \\
                \expandafter\colvecnext
        \else
                \end{pmatrix}
        \fi
}

\begin{document}

\title{Strong quantum fluctuations from competition between magnetic phases in a pyrochlore iridate}

\author{Henrik Jacobsen}
\altaffiliation{Present address: Laboratory for Neutron Scattering, Paul Scherrer Institute, CH-5232 Villigen PSI, Switzerland}
\affiliation{Department of Physics, Oxford University, Clarendon Laboratory, Oxford, OX1 3PU, United Kingdom}
\author{Cameron D. Dashwood}
\affiliation{London Centre for Nanotechnology and Department of Physics and Astronomy, University College London, London WC1E 6BT, United Kingdom}
\author{Elsa Lhotel}
\affiliation{Institut N\'eel, Centre National de la Recherche Scientifique and Universit\'e Joseph Fourier, BP 166, 38042 Grenoble Cedex 9, France}
\author{Dmitry Khalyavin}
\affiliation{ISIS Pulsed Neutron and Muon Source, STFC Rutherford Appleton Laboratory, Harwell Campus, Didcot, OX11 0QX, United Kingdom}
\author{Pascal Manuel}
\affiliation{ISIS Pulsed Neutron and Muon Source, STFC Rutherford Appleton Laboratory, Harwell Campus, Didcot, OX11 0QX, United Kingdom}
\author{Ross Stewart}
\affiliation{ISIS Pulsed Neutron and Muon Source, STFC Rutherford Appleton Laboratory, Harwell Campus, Didcot, OX11 0QX, United Kingdom}
\author{Dharmalingam Prabhakaran}
\affiliation{Department of Physics, Oxford University, Clarendon Laboratory, Oxford, OX1 3PU, United Kingdom}

\author{Desmond F. McMorrow}
\affiliation{London Centre for Nanotechnology and Department of Physics and Astronomy, University College London, London WC1E 6BT, United Kingdom}

\author{ Andrew. T. Boothroyd }
\affiliation{Department of Physics, Oxford University, Clarendon Laboratory, Oxford, OX1 3PU, United Kingdom}

\date{\today}

\begin{abstract}
We report neutron diffraction measurements of the magnetic structures in two pyrochlore iridates, \Ybtts{} and \Lutts{}. Both samples exhibit the all-in-all-out magnetic structure  on the Ir$^{4+}$ sites below $T_{\rm N} \simeq 150$\,K, with a low temperature moment of around 0.45\,$\mu_{\rm B}$/Ir.  Below 2\,K, the Yb moments in \Ybtts{} begin to order ferromagnetically. However, even at 40\,mK the ordered moment is only 0.57(2)\,$\mu_{\rm B}$/Yb, well below the saturated moment of the ground state doublet of Yb$^{3+}$ (1.9\,$\mu_{\rm B}$/Yb), deduced from magnetization measurements and from a refined model of the crystal field environment, and also significantly smaller than the ordered moment of Yb in Yb$_2$Ti$_2$O$_7$ (0.9\,$\mu_{\rm B}$/Yb). A mean-field analysis shows that the reduced moment on Yb is a consequence of enhanced phase competition caused by coupling to the all-in-all-out magnetic order on the Ir sublattice.
\end{abstract}
\maketitle

\section{Introduction}
The extended family of pyrochlore oxides $A_2B_2$O$_7$ exhibits an enormous range of exotic and interesting magnetic phenomena\cite{Gardner2010}. This  richness of behavior stems from the structure of the $A$ and $B$ sublattices, which form interpenetrating nets of corner-sharing tetrahedra, and from the local anisotropy of the magnetic ions. For example, strong Ising anisotropy leads to classical spin ice behaviour and emergent magnetic monopoles, as found in Dy$_2$Ti$_2$O$_7$ and Ho$_2$Ti$_2$O$_7$,\cite{Harris1997,Castelnovo2008,Morris2009,Fennell2009} whereas XY anisotropy leads to unconventional ordered states in Yb and Er based compounds \cite{Ross2011,Chang2012b,Champion03,Petit17}.

The magnetic ground state in these materials can be tuned by control parameters such as pressure (mechanical and chemical) \cite{Snyder02, Mirebeau02, Takatsu16, Kermarrec2017, Shirai17} and external magnetic fields \cite{Tabata06, Ruff08, Scheie17, Thompson2017}. An alternative approach, however, is to create an effective internal magnetic field at the $A$ site by substitution of a magnetic ion on the $B$ site. It is then possible that the staggered field generated by a magnetic coupling between the two sites can produce competition between their respective preferred ground states \cite{Lefrancois2017}.

\begin{figure}
\includegraphics[width=0.5\textwidth]{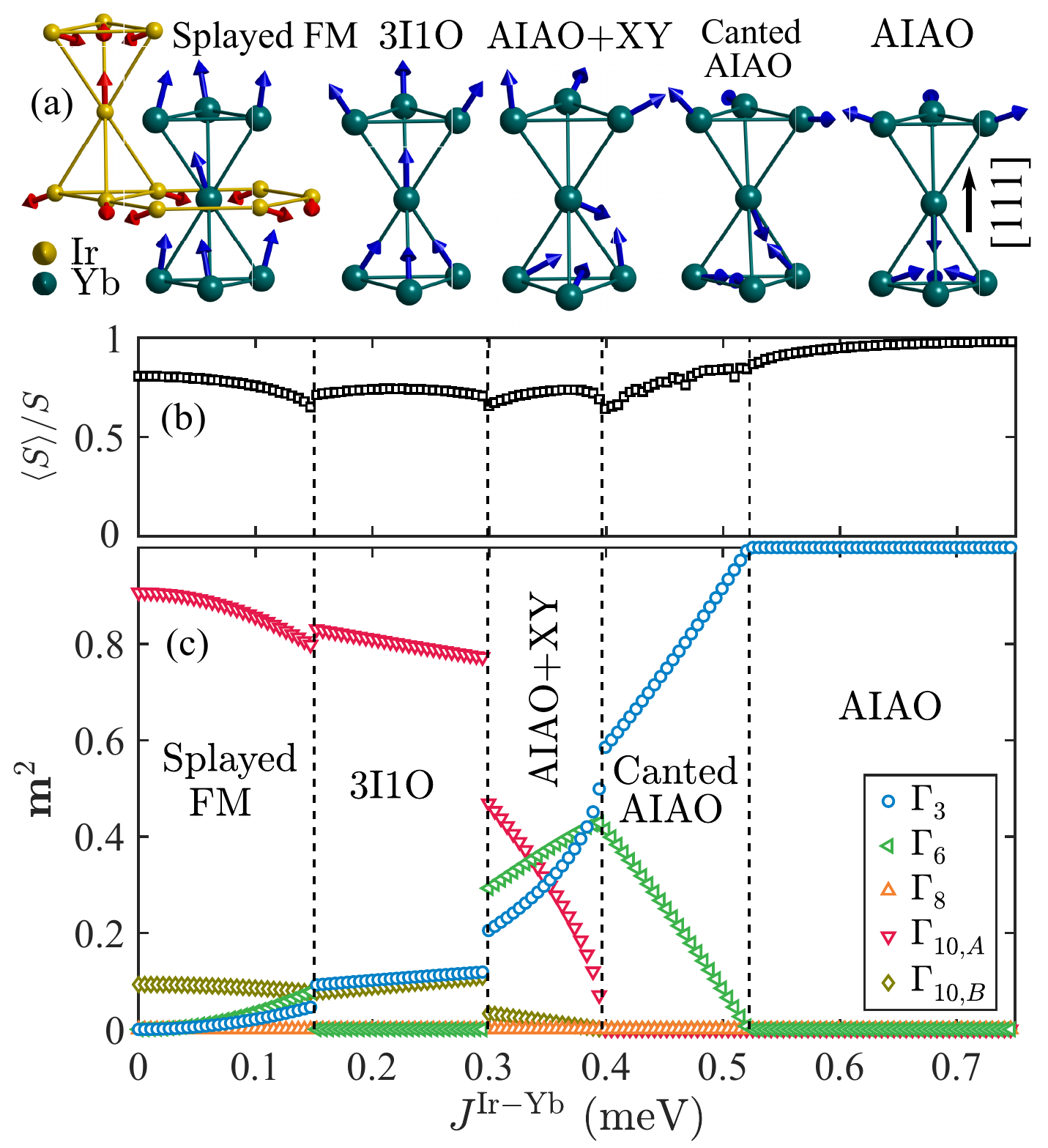}
\caption{Magnetic order of \Ybtts{} as a function of the strength of the coupling between Ir and Yb ($J^{\rm Ir-Yb}$). 
(a) Magnetic structures of the five phases predicted in our model. The Ir moments (short red arrows) point either all-in or all-out (AIAO) relative to the center of the tetrahedra. Depending on $J^{\rm Ir-Yb}$, the order of the Yb moments (long blue arrows) is either (i) a splayed ferromagnet, as found in \YbTitts{}, (ii) three-in one-out (3I1O), (iii) a combination of all-in-all-out and XY antiferromagnet (AIAO$+$XY), (iv) canted AIAO, or (v) AIAO. (b) The calculated ordered moment as function of the coupling strength, showing a reduction close to the phase boundaries.
(c) The phase diagram showing the square of the order parameter for each of the five irreps. Details are given in the text.
}
\label{fig:Fig1}
\end{figure}

A system in which such a scenario might occur is the iridate pyrochlore oxides $A_2$Ir$_2$O$_7$, where $A$ is a trivalent  lanthanide.  The magnetic properties of the iridate pyrochlores have been quite extensively studied, and the Ir sites have been found to develop long-range order at a temperature  $T_{\rm N}$ close to the onset of a metal-to-insulator transition, where $T_{\rm N} = 115$--150\,K for $A = $ Sm--Lu, and $T_{\rm N} \simeq 33$\,K for $A =$ Nd \cite{Taira2001,Matsuhira2011,Disseler2012a,Ishikawa2012,Witczak-krempa2014}. The exception is Pr$_2$Ir$_2$O$_7$, which exhibits no conventional magnetic order down to 70\,mK.\cite{Machida2007} In the $A_2$Ir$_2$O$_7$ compounds studied so far, the ordered magnetic moments on the Ir sites are small ($\sim 0.5$\,$\mu_{\rm B}$) and point either all towards the center of the tetrahedron (the local [111] direction) or directly away from it --- the so-called all-in-all-out (AIAO) structure, see Fig.~\ref{fig:Fig1}(a)  \cite{Tomiyasu2012,Sagayama2013,Disseler2014,Donnerer2016,Guo2016,Guo2017,Lefrancois2015,Yang2017}.

If the magnetic coupling between the Ir and the $A$ sites is strong then the Ir order will induce AIAO order on the $A$ site, as observed in e.g.~\Ndtts{} (Ref.~\onlinecite{Guo2016}) and \Tbtts{} (Ref.~\onlinecite{Guo2017}). 

In this work we investigate the effect of the staggered molecular field from Ir in \Ybtts{}, using a combination of neutron diffraction, neutron spectroscopy and macroscopic measurements. We find that competition between the planar single-ion anisotropy of Yb$^{3+}$, the splayed ferromagnetism favored by the Yb--Yb coupling, and the AIAO order favored by coupling to the ordered Ir spins ($J^\text{Ir-Yb}$), tends to suppress magnetic order of the Yb moments and leads to strong quantum fluctuations down to the lowest temperatures. We show how these results can be understood in terms of competing phases induced by $J^\text{Ir-Yb}$.

\section{Experimental details}

We prepared polycrystalline samples of \Ybtts{} and \Lutts{} by the conventional solid-state reaction method using high purity (99.99\%) Yb$_2$O$_3$/Lu$_2$O$_3$ and IrO$_2$ powders. \cite{Prabhakaran2017}. 
Standard magnetisation, susceptibility and resistivity measurements were carried out using a Quantum Design Magnetic Properties Measurement System (MPMS) and a Physical Properties Measurement System (PPMS) between 2 and 300\,K. Magnetization and AC susceptibility measurements between 0.1 and 4.2\,K were carried out in a SQUID magnetometer equipped with a dilution fridge developed at the Institut N\'eel. 
Specific heat measurements were performed on a PPMS equipped with a He3 pumping system. Measurements were successively performed on two \Ybtts{} pellets of 6.19 and 9.85 mg, and on a 5.89 mg \Lutts{} pellet to subtract lattice and iridium contributions.

Heat capacity measurements were carried out for temperatures down to 0.5 K.
Neutron diffraction measurements were performed at the WISH diffractometer\cite{Chapon2011,WISH_data} at the ISIS Facility. 
 
Samples of polycrystalline \Lutts{} and  \Ybtts{} with masses 3.4\,g and 5.0\,g, respectively,  were loaded into aluminium cans which had the form of a cylindrical annulus with an average radius of 12\,mm and radial spacing 1\,mm. The cans were installed in a standard helium cryostat for measurements at temperatures down to 1.5\,K. 
Both samples were measured at 1.5\,K and 160\,K for approximately 4 hours, and the \Ybtts{} sample was additionally measured at 60\,K for 4 hours and in shorter runs ($\sim30$\,min) at several other temperatures between 20\,K and 140\,K. 
A second experiment, with the sample of \Ybtts{} now in a copper can of similar annular geometry installed in a dilution fridge, was carried out at several temperatures between 40\,mK and 900\,mK, and at 10\,K. The 40 and 900\,mK runs were measured for 3 hours, the 10\,K run for 1.5 hours, and the intermediate temperatures between 40\,mK and 900\,mK for 1 hour each. The sample was kept at 40\,mK for 24 hours before the experiment started. The diffraction data were reduced using MANTID \cite{Arnold2014}, and structural refinements were performed with FullProf\cite{Rodriguez-Carvajal1993}.

 Neutron inelastic scattering measurements were carried out at the ISIS Facility on the  MAPS time-of-flight spectrometer\cite{Perring1994,Ewings2019,MAPS_data}.
 We measured the same polycrystalline \Ybtts{} sample as used in the neutron diffraction experiments.  The sample was spread as evenly as possible inside a $4 \times 13$\,cm$^2$ aluminium sachet which was inserted in an aluminium cylindrical can. The can was then mounted in a closed-cycle refrigerator and cooled to a base temperature of 5.5\,K. Spectra were recorded with neutrons of incident energy $E_{\rm i}=200$\,meV and $E_{\rm i}=110$\,meV for 13.5 and 9 hours, respectively.

 The spectra were normalized to vanadium and corrected for sample absorption assuming an evenly loaded sample, and for the magnetic form factor of Yb$^{3+}$, $f^2(Q)$, as well as for a small offset on the energy axis.  An accurate absolute calibration proved not to be possible because of the large neutron absorption cross-section of Ir. The appendix includes more details of the experiment and analysis.

\section{Results}

\subsection{Single-ion magnetism of Yb}

\begin{figure}
\includegraphics[width=0.48\textwidth]{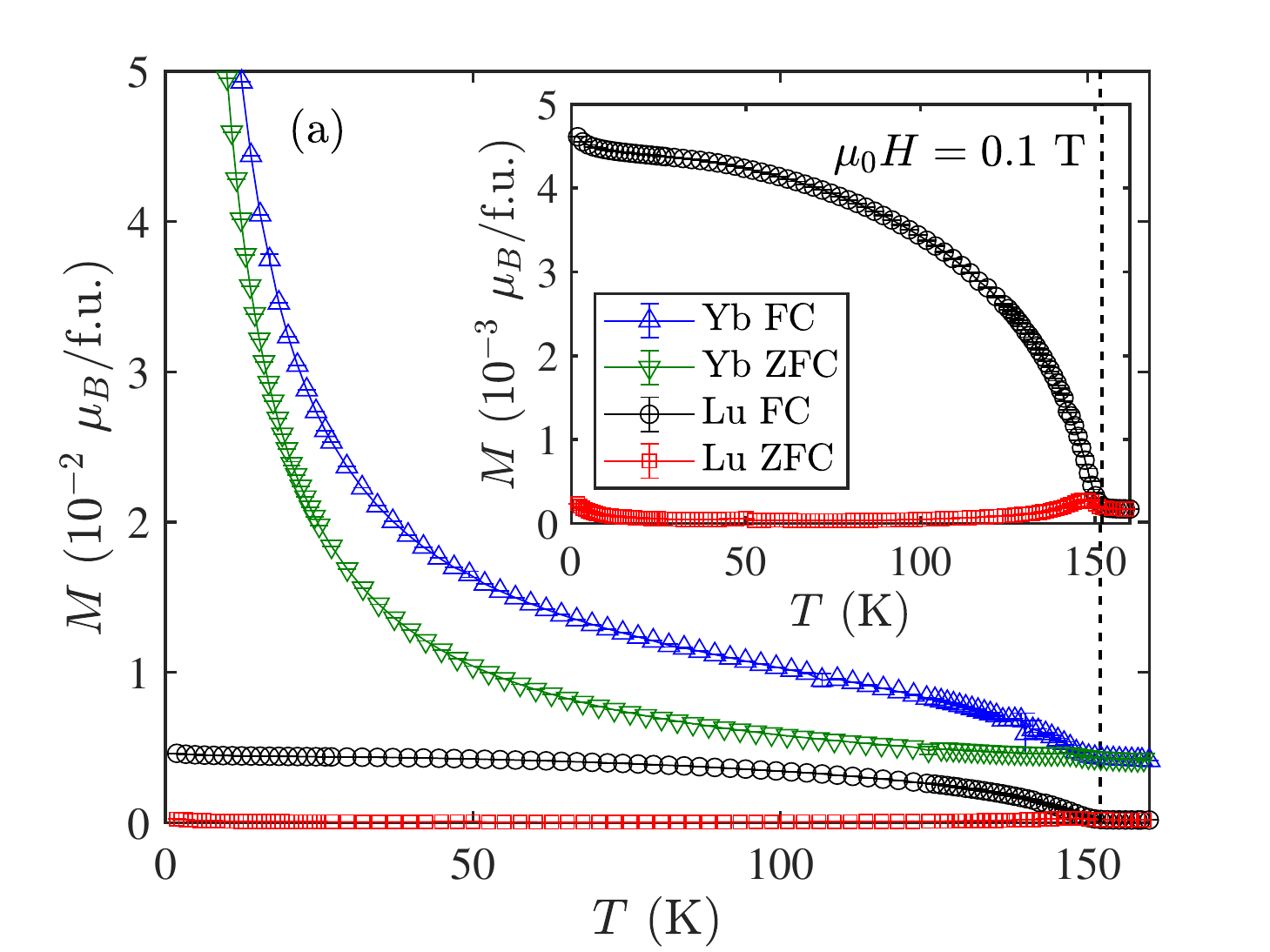}
\includegraphics[height=0.14\textheight]{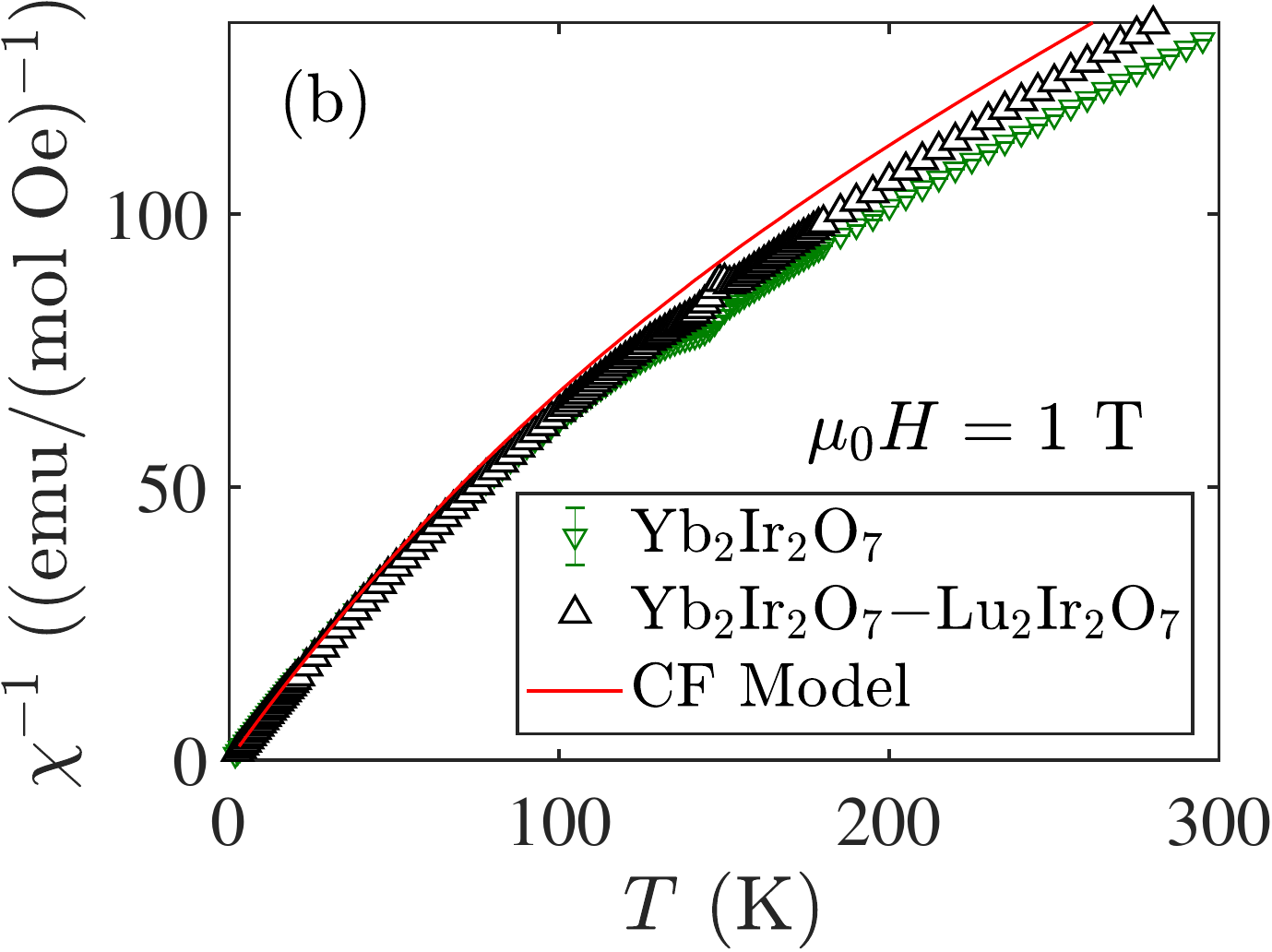}
\includegraphics[height=0.14\textheight]{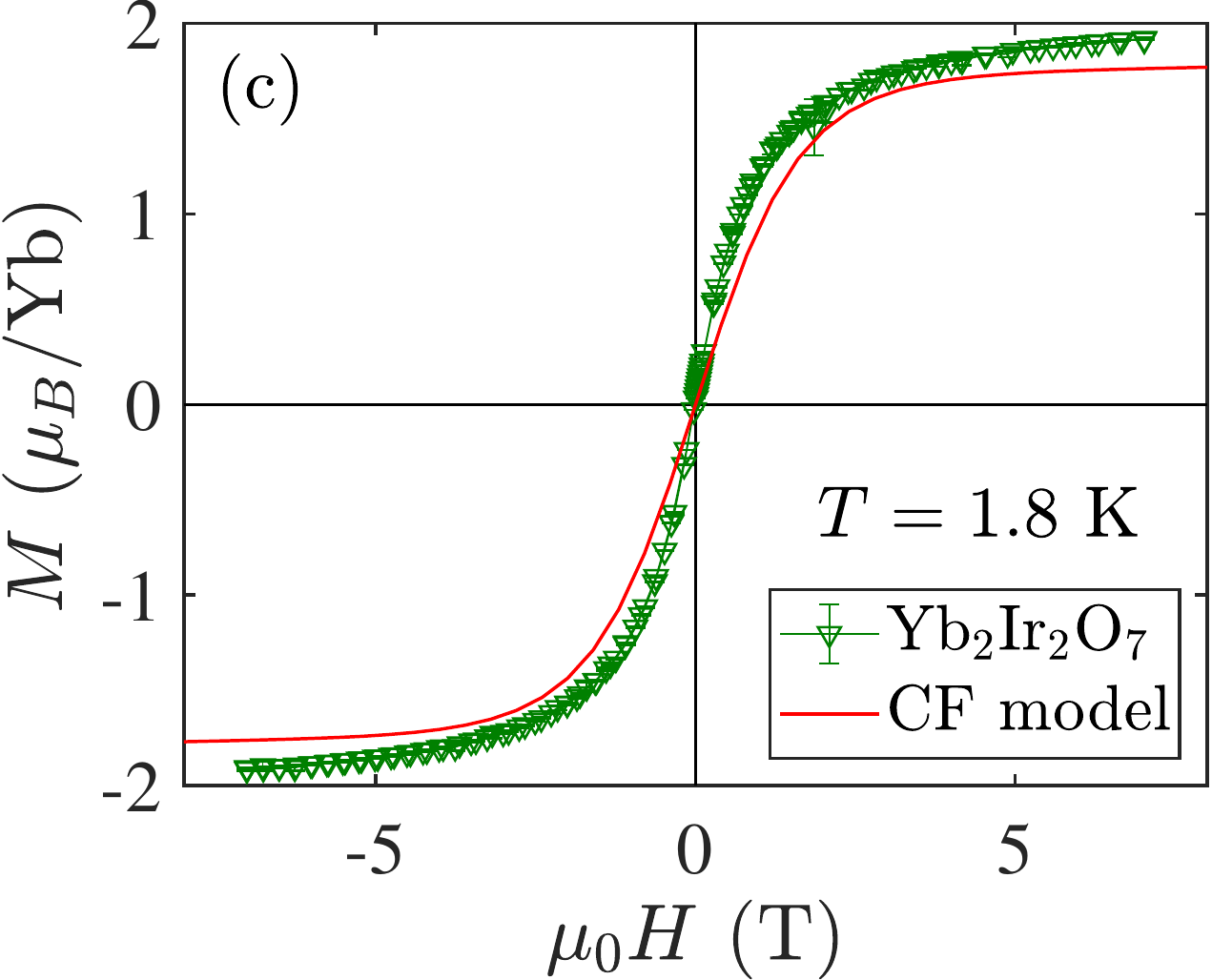}
 \caption{(a) Magnetization of \Lutts{} and \Ybtts{} as function of temperature measured in an applied field of 0.1\,T. The insert shows the \Lutts{} data on an enlarged vertical scale.
 (b) Magnetic susceptibility of powdered \Ybtts{} as function of temperature in an applied field of $\mu_0H=1$\,T. We show both the raw data and data corrected for the measured susceptibility of  \Lutts{} to remove the contribution from Ir. The red line shows the calculated susceptibility based on our single-ion model for Yb$^{3+}$.
 (c) the magnetization of  \Ybtts{} at $T=1.8$\,K as a function of applied field, together with the magnetization of Yb$^{3+}$ calculated from our single-ion model.
}
 \label{fig:Fig2}
\end{figure}

The magnetization curves as a function of temperature for both samples (Fig.~\ref{fig:Fig2}(a)) show  bifurcations between field-cooled and zero-field-cooled curves at $T_{\rm N} \simeq 150$\,K, indicating the onset of magnetic ordering of the Ir sublattice. In \Ybtts{} there is a large additional paramagnetic signal from the Yb$^{3+}$ magnetic moments.

Figure~\ref{fig:Fig2}(b) shows the measured inverse susceptibility (calculated as $M/H$) for the \Ybtts{} sample. We show both the raw data, and corrected data after subtraction of the measured susceptibility of \Lutts{} as an estimate of the magnetic contribution from iridium. The red line shows the susceptibility calculated from our model for the crystal field detailed below. 

The magnetization of \Ybtts{} as a function of applied field is shown in Fig.~\ref{fig:Fig2}(c), along with the calculation from our crystal field model. The saturated magnetization at $T=1.8$\,K is 1.9\,$\mu_{\rm B}$/Yb.      

\begin{figure}
\centering
    \includegraphics[width=0.48\textwidth]{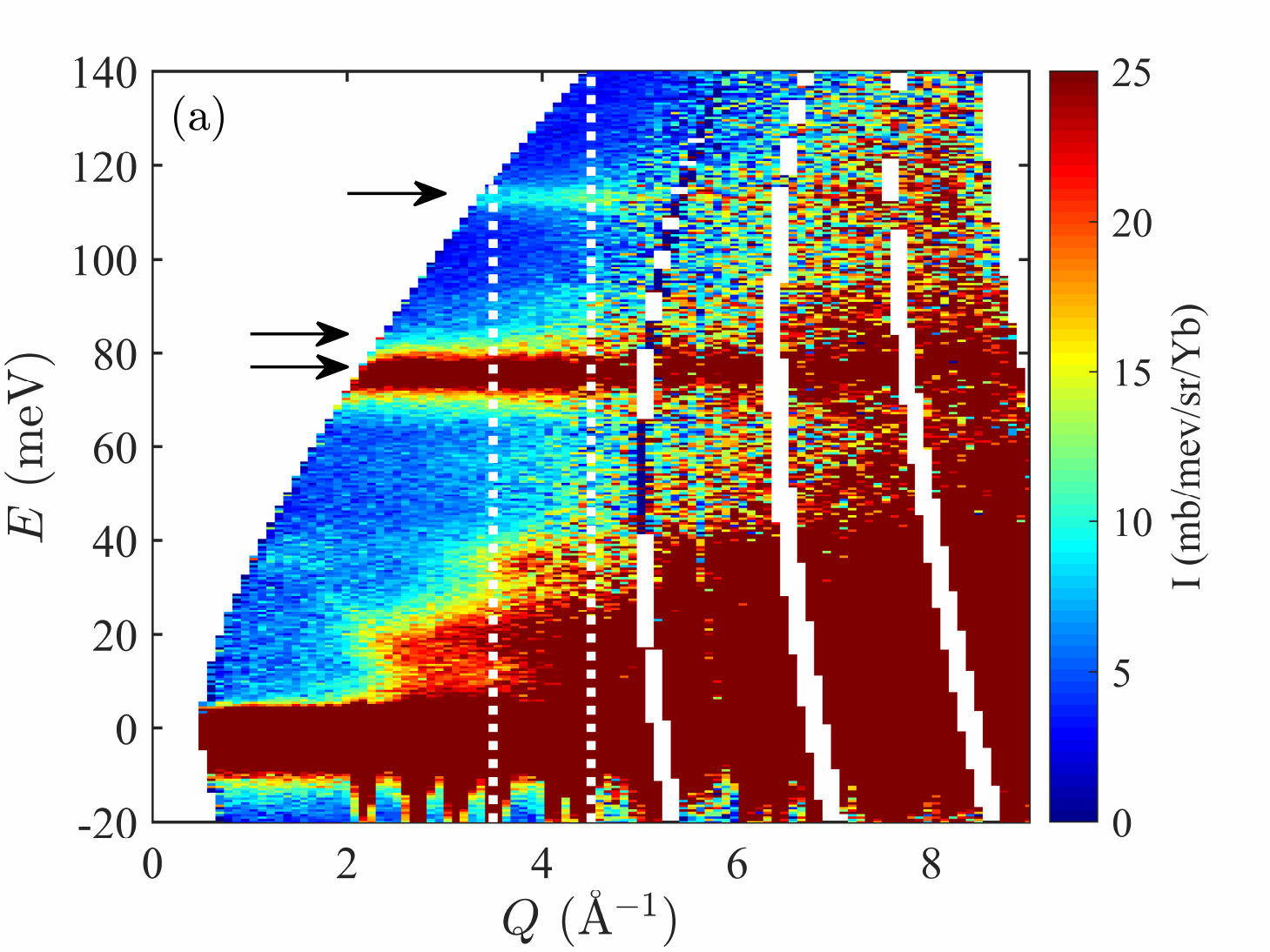}
        \includegraphics[width=0.48\textwidth]{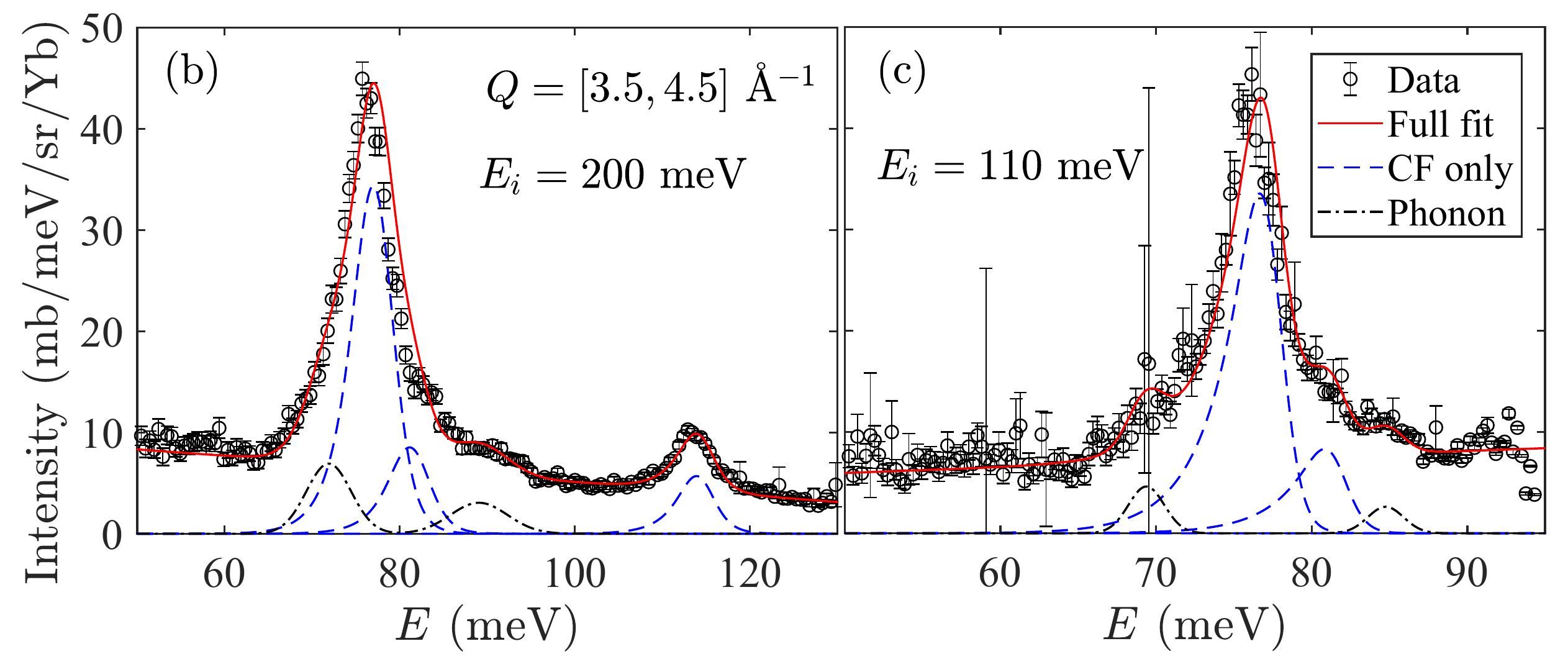}
    \caption{Crystal field excitations of Yb$^{3+}$ in \Ybtts{} measured on MAPS with an incoming energy of 200\,meV. The data have been  normalised to mbarn/meV/sr/Yb using a vanadium standard and corrected for the absorption of iridium and the magnetic form factor of Yb$^{3+}$ as described in the text. (a) shows the intensity as function of energy transfer, $E$, and scattering vector, $Q$ with arrows indicating the crystal field excitations. (b) A constant $Q$ cut through the data shown in (a) within the limits shown by the dotted white lines in (a). The red line shows the full fit to the data consisting of the crystal field excitations (dotted blue line), two phonons (solid black line) and a sloping background (not shown).  (c) Same as (b) but for $E_i=110$\,meV.}
    \label{fig:Fig3}
\end{figure}

In order to interpret the single-ion magnetic response of Yb we have measured the spectrum of crystal-field excitations within the $^2F_{7/2}$ term of Yb$^{3+}$ ($4f^{13}$) in our \Ybtts{} sample by inelastic neutron scattering. Figure~\ref{fig:Fig3}(a) is a color map of the corrected intensity as function of scattering vector, $Q$, and energy transfer, $E$. We made a constant-$Q$ cut through the data, averaging the intensity over $3.5<Q<4.5$\,\AA{}$^{-1}$. These cuts are shown in Figs.~\ref{fig:Fig3}(b) and (c) for $E_i=200$ and $110$\,meV, respectively. 

We identify two clear peaks from crystal field excitations at 76.6(6) and 113.5(3)\,meV. The peak at 76.6\,meV has shoulders on both sides which can be modelled with peaks centred near 71 and 81\,meV. The 81 meV peak is attributed to a crystal field excitation, while the 71 meV peak is a phonon. The crystal field excitations are marked with arrows in Fig.~\ref{fig:Fig3}(a). Details of the analysis can be found in the Appendix.

The crystal field excitations in the pyrochlores can be modeled with the Hamiltonian
\begin{align}
    \mathcal{H}= &
   \, B_0^2 C_0^{(2)}
    + B_0^4 C_0^{(4)}
    + B_3^4 C_3^{(4)}\nonumber\\
    & + B_0^6 C_0^{(6)}
    + B_3^6 C_3^{(6)}
    + B_6^6 C_6^{(6)},
    \label{eq:Eq1}
\end{align}
where $B_q^k$ are numerical coefficients and $C_q^{(k)}$ are the Wybourne tensor operators, given by
\begin{align}
    C_q^{(k)}(\theta,\phi) = \sqrt{\frac{4\pi}{2k+1}}Y_{k,q}(\theta,\phi),
\end{align}
with $Y_{k,q}$ the spherical harmonic functions (here we use the Condon-Shortley phase convention). We use the SPECTRE program for all crystal field calculations\cite{Spectre}. We note that unlike the commonly used Stevens formalism, SPECTRE can include all states of the $4f^n$ configuration, which for Yb$^{3+}$ means both the $J=7/2$ and the $J=5/2$ levels. 
We ignore the Yb--Yb and Yb--Ir exchange interactions as they are much smaller than the crystal field potential, and have negligible effect on the crystal field spectrum. 

We find good agreement between the data and our model, see Figs.~\ref{fig:Fig3}(b) and (c). The susceptibility and magnetisation calculated with this model are shown in Fig.~\ref{fig:Fig2}. The model  slightly underestimates the measured susceptibility and magnetisation. The  discrepancy could be an effect of the exchange interactions which our single-ion model does not take into account. 

The Yb$^{3+}$ ions in \Ybtts{} have weak planar single-ion anisotropy from the crystal field. We find that  the $g$ tensor components parallel and perpendicular to the local $<$111$>$ axes are $g_\parallel = 2.3$ and $g_\perp = 4.0$, and the saturated moment is $M_s=1.8$\,$\mu_\textrm{B}$/Yb, slightly smaller than the value 1.9 found from magnetisation. 

\subsection{Crystal structure refinement}
\begin{figure}
\centering
\includegraphics[width=0.48 \textwidth]{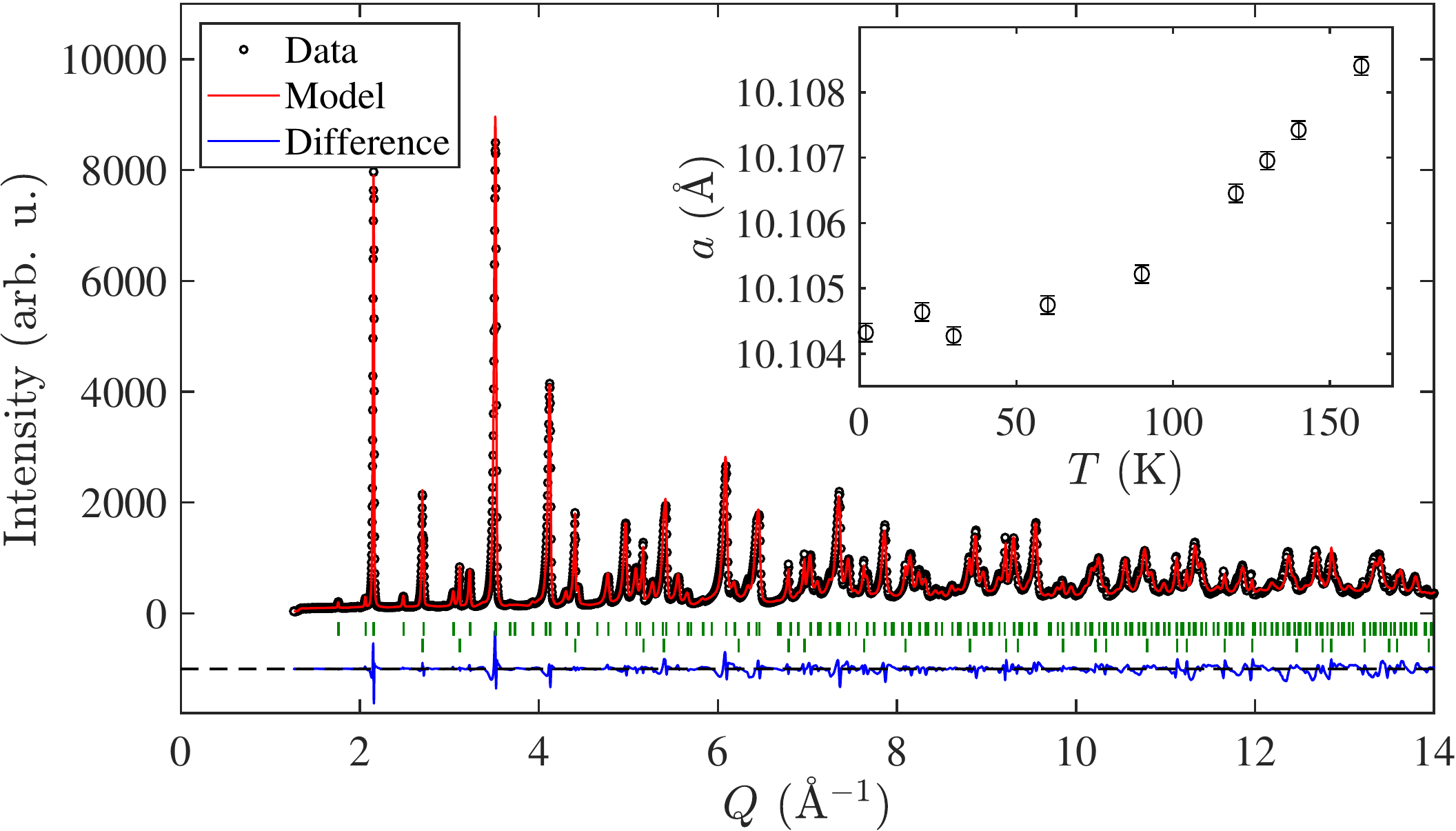}
\caption{Neutron powder diffraction pattern of \Ybtts{} measured at 1.5\,K. The red line through the data is a Rietveld refinement of the crystal structure together with a small signal from the aluminium sample holder. The tick marks shows the peaks positions of these two contributions, and the line underneath is the difference $I_{\rm model}-I_{\rm obs}$. The intensity of the magnetic scattering is too weak to include a magnetic structure model in the refinement. The inset shows the refined lattice constant as function of temperature.
}
\label{fig:Fig4}
\end{figure}

 Figure~\ref{fig:Fig4} shows neutron diffraction data on \Ybtts{} recorded at 1.5\,K in banks 5 and 6 of the Wish diffractometer, along with a Rietveld refinement of the crystal structure. A small amount of parasitic scattering from the aluminium sample container was also included in the refinement.   The magnetic signal from the Ir$^{4+}$ moments is weak and cannot be refined without background subtraction, and has thus been excluded from the nuclear refinement. 

Both samples were found to be $>98\%$ pure, and their structures were refined in the space group $Fd\bar{3}m$ with lattice constant $a=10.094(1)$\,\AA{} (\Lutts{}) and $a=10.104(1)$\,\AA{} (\Ybtts{}) at 1.5\,K.

 We find excellent agreement between our data and the refinement. The Yb$^{3+}$ ions occupy the 16d Wyckoff positions at $0.5,0.5,0.5$; the Ir$^{4+}$ ions occupy the 16c Wyckoff positions at $0,0,0$; and the O$^{2-}$ ions occupy the Wyckoff positions 8b at $0.375,0.375,0.375$ and 48f at $x,0.125,0.125$. At 1.5\,K we find $x=0.340(1)$. The Bragg R-factor of the refinement is 3.17.

The refined oxygen content is 6.97(4). 
Site-mixing could not be refined due to the large neutron absorption of Ir and the fact that the neutron- and x-ray scattering cross sections for Yb and Ir are quite similar, but as the samples are single-phase and near-stoichiometric in oxygen, any site mixing is likely to be minimal.

The large neutron absorption cross-section of Ir is difficult to correct for in our data. We find the fitting routine in some refinements compensates for the absorption by giving unphysical values of the isotropic displacement parameters, $B_\text{iso}$. To investigate the significance of this, we have repeated the refinements for various fixed values of $B_\text{iso}$. We find that the refined magnetic parameters presented below are virtually insensitive to $B_\text{iso}$. This is because the magnetic parameters are essentially determined by the ratio of the intensity of the nuclear Bragg peaks to that of the magnetic Bragg peaks, and these intensities are impacted equally by the absorption

\subsection{Magnetic structures in Lu$_2$Ir$_2$O$_7$ and Yb$_2$Ir$_2$O$_7$}

\begin{figure}
\includegraphics[width=0.48\textwidth]{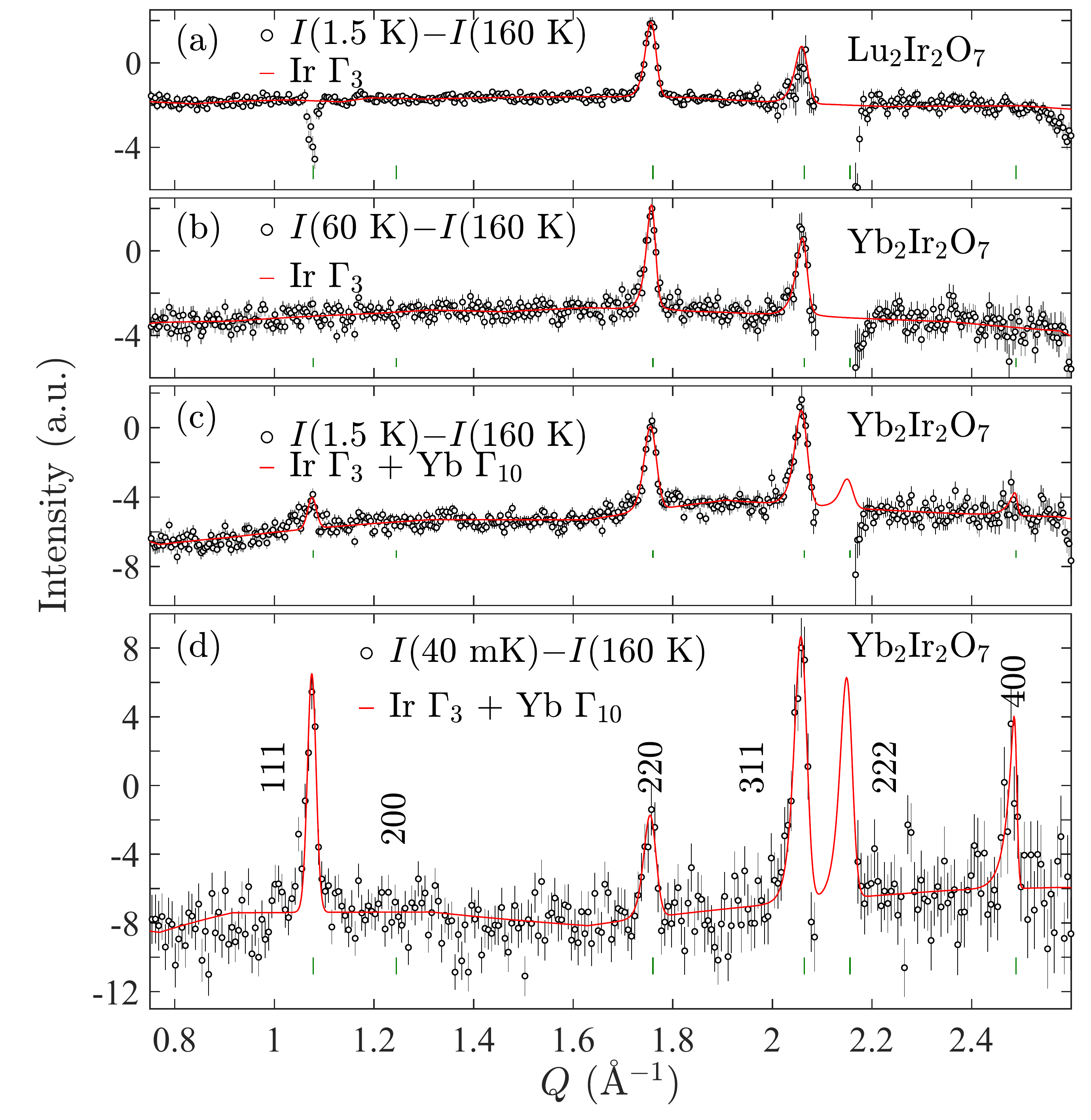}
\caption{ 
Neutron powder diffraction from (a) \Lutts{} at  2\,K, and (b)--(d) \Ybtts{} at 60\,K, 1.5\,K and 40\,mK, respectively.  Data measured in the paramagnetic phase at 160\,K have been subtracted to isolate the magnetic signal, with allowance for thermal expansion of the lattice parameters (see the Appendix for details). Data near the strong 222 nuclear reflection have been masked due to an imperfect subtration.  The red lines in (a) and (b) are the result of Rietveld refinement of an AIAO magnetic structure on the Ir sublattice. In (c) and (d), an additional ferromagnetic component on the Yb sublattice has been refined.
}
\label{fig:Fig5}
\end{figure}

Figure~\ref{fig:Fig5}(a) shows neutron diffraction from \Lutts{} at 1.5\,K as a function of scattering vector $Q$, recorded in banks 2 and 9 of WISH. Data recorded in the paramagnetic phase at 160\,K have been subtracted, and the data have been rebinned for clarity.  
To improve subtraction we scaled the time-of-flight of the 160\,K data by the ratio of lattice constants at 160\,K and the other temperature. 

We observe magnetic Bragg peaks corresponding to the 220 and 311 reflections, but there are no detectable peaks at the positions of the 111, 200 and 400 reflections. A small negative signal at $Q = 1.08$\,\AA$^{-1}$ is due to an incomplete subtraction of the 111 nuclear reflection.  The magnetic peak positions are consistent with a $\kkk=0$ magnetic structure. 

The magnetic moments on the Yb/Lu and Ir sites form a basis for a representation of the space group $Fd\bar{3}m$ with propagation vector $\kkk=0$ which decomposes as\cite{Guo2017,basireps}
\begin{align}
\Gamma_\text{mag} = \Gamma_3^1+\Gamma_6^2+\Gamma_8^3 + 2\Gamma_{10}^3,
\end{align}
Here, $\Gamma_3^1$ is the irreducible representation (irrep) which describes the AIAO structure, $\Gamma_6^2$ contains the $\psi_2$ and $\psi_3$ structures \cite{Champion03}, $\Gamma_8^3$ is the Palmer-Chalker state \cite{Palmer2000}, and $\Gamma_{10,A}^3$ and $\Gamma_{10,B}^3$ are a pure ferromagnet and a non-collinear ferromagnet, respectively.  From now on, we omit the superscript which indicates the dimension of the irrep. Only the AIAO structure ($\Gamma_3$) is consistent with our data for \Lutts{} at 1.5\,K. The fit to this magnetic structure is shown in Fig.~\ref{fig:Fig5}(a) and is in good agreement with the data.

The AIAO structure also gives a very good description of our data on \Ybtts{} for temperatures down to 20\,K [Fig.~\ref{fig:Fig5}(b)]. We conclude, therefore, that the Ir sublattice in \Lutts{} and \Ybtts{} orders in the same AIAO structure as found in other $A_2$Ir$_2$O$_7$ iridates\cite{Tomiyasu2012,Sagayama2013,Disseler2014,Donnerer2016,Guo2016,Guo2017,Lefrancois2015,Yang2017}. Note that in \Ybtts{}, the nuclear 111 peak is much weaker than in \Lutts{}, and the subtraction is therefore near perfect for \Ybtts{}.

\begin{figure}
 \includegraphics[width=0.4\textwidth]{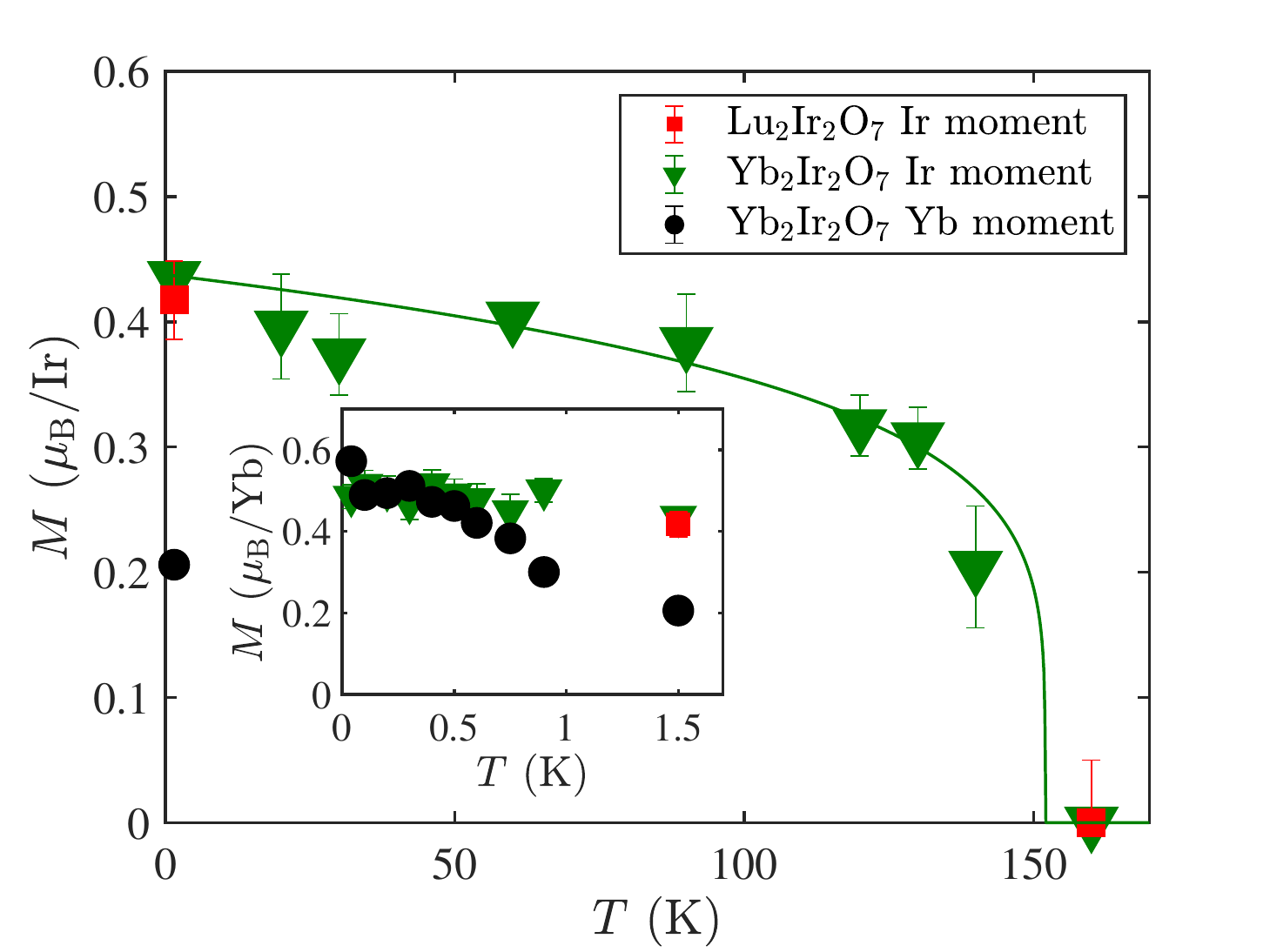}
 \caption{
Main panel: Temperature dependence of the refined magnetic moment on Ir in \Lutts{} and \Ybtts{}, assuming only AIAO order. The line is to guide the eye. Insert: Low temperature region showing also the moment on the Yb sublattice assuming only a ferromagnetic component.  Data below 1\,K were measured in a dilution fridge.}
\label{fig:Fig6}
\end{figure}

The refined magnetic moment on the Ir sites is shown as function of temperature in Fig.~\ref{fig:Fig6}. It rises rapidly upon cooling below 150\,K, reaching 0.45(2)\,$\mu_{\rm B}$ in \Lutts{}, and 0.44(1)\,$\mu_{\rm B}$ in \Ybtts{} at low temperatures, similar in magnitude to other iridates such as Gd$_2$Ir$_2$O$_7$ [0.30(3)\,$\mu_{\rm B}$]\cite{Lefrancois2019}, \Ndtts{} [0.34(1)\,$\mu_{\rm B}$]\cite{Guo2016} and \Tbtts{} [0.55(3)\,$\mu_{\rm B}$]\cite{Guo2017}. In perfect octahedral symmetry, the $5d^5$ configuration of Ir$^{4+}$ combined with spin--orbit coupling forms a $J_\text{eff}=\frac{1}{2}$ ground state with a magnetic moment $\langle \mu \rangle = \frac{1}{3}\mu_{\rm B}$ (Ref.~\onlinecite{Kim2008}).  
The trigonal distortion mixes the $t_{2g}$ and $e_g$ levels, making the $J_\text{eff}=\frac{1}{2}$ picture only approximately correct and leading to an observed moment which is larger than $\frac{1}{3}\mu_{\rm B}$ (Refs.~\onlinecite{Stamokostas2018,Shinaoka2015,Zhang2017}).

We now turn to the Yb sublattice. Figures~\ref{fig:Fig5}(c) and (d) display neutron diffraction data on \Ybtts{} at 1.5\,K and 40\,mK, respectively.  At 1.5\,K, there is some enhancement of intensity at the positions of the 111 and 400 reflections, both of which grow into strong peaks upon cooling to 40\,mK. These peaks indicate ordering of the Yb$^{3+}$ sublattice in a different structure than the AIAO ordering of the Ir moments. The magnetic peaks below 1\,K are slightly broader than the resolution function and gradually sharpen upon cooling, indicating correlation lengths of $\xi=1/\Gamma\approx 100-300$\,\AA{}, where $\Gamma$ is the half width at half maximum of the peaks.

We find excellent agreement between our data and a model in which only the $\Gamma_{10,A}$ component is refined, indicating ferromagnetic ordering as also found in \YbTitts{}.\cite{Yasui2003,Chang2012b,Gaudet2016,Yaouanc2016,Pecanha-antonio2017} We have tried including the other irreps in the refinement, but find negligible components of $\Gamma_6$ [0.04(4)\,$\mu_\text{B}$], $\Gamma_8$ [0.04(4)\,$\mu_\text{B}$] and $\Gamma_{10,B}$ [0.04(4)\,$\mu_\text{B}$]. These values constrain any splay angle to be less than about $10^\circ$. 

The 220 reflection is absent in the pure $\Gamma_{10,A}$ ferromagnetic structure, but we observe a slight enhancement of the 220 peak at 40\,mK relative to 1.5\,K. This enhancement could be caused either by (a) a small increase in the Ir AIAO moment, or (b) a small AIAO component (around 10\% of the total moment) of the Yb moments. The quality of our diffraction data is insufficient to distinguish these possibilities, but our calculations presented below indicate that scenario (b) is the most likely.

The temperature dependence of the refined ordered  moment on Yb is shown in the insert to Fig.~\ref{fig:Fig6}. The two models, (a) and (b), give very similar ordered moments on the Yb site, which we find to be $0.57(2)$\,$\mu_{\rm B}$ at 40\,mK. This moment is much smaller than the saturated moment of the Kramers doublet ground state of Yb$^{3+}$, that we determined to be $M_{\rm s} \simeq 1.9$\,$\mu_{\rm B}$  from our low temperature bulk magnetometry and  crystal field model refined against neutron spectroscopy data presented above.

\begin{figure}
\centering
 \includegraphics[width=0.4\textwidth]{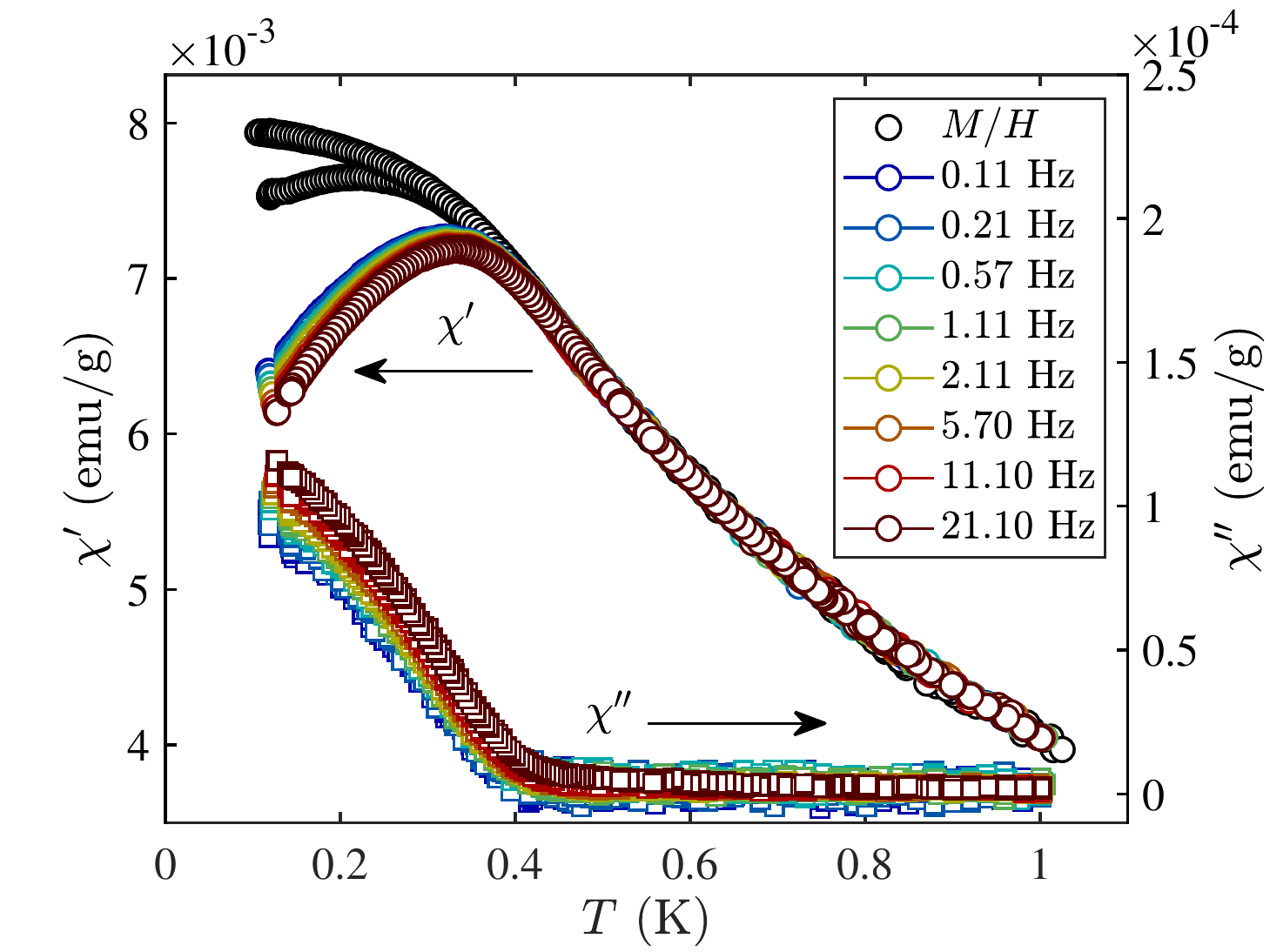}
 \caption{The real $(\chi')$ and imaginary ($\chi''$) parts  of the measured AC susceptibility signal from \Ybtts{}, showing that spin freezing sets in at $T\simeq 0.4$\,K.}
\label{fig:Fig7}
\end{figure}

The existence of magnetic order over long distances in which only a small fraction of the total moment is ordered suggests that there are significant low frequency fluctuations. Further evidence in support of this is found in the AC susceptibility  measurements presented in Fig.~\ref{fig:Fig7} for frequencies between 0.11 and 21.1\,Hz. There is a broad peak in the real part of the AC susceptibility $\chi'(\omega)$ at around 350\,mK, and the imaginary part $\chi''(\omega)$ becomes non-zero at about 400\,mK. A bifurcation in DC susceptibility is also seen at 350\,mK. These observations may indicate the onset of magnetic order at this temperature. 
The peak in $\chi'(\omega)$ is broader and rounder than what would be expected for a continuous phase transition, and has a slight frequency dependence that may be associated with domain wall dynamics. Our results are quite different from the first-order ferromagnetic transition in \YbTitts{}, which exhibits a sharp cusp in $\chi'(\omega)$ and a rapid rise in $\chi''(\omega)$ at the phase transition\cite{Lhotel2014}. 

\begin{figure}
\centering
\includegraphics[width=0.4\textwidth]{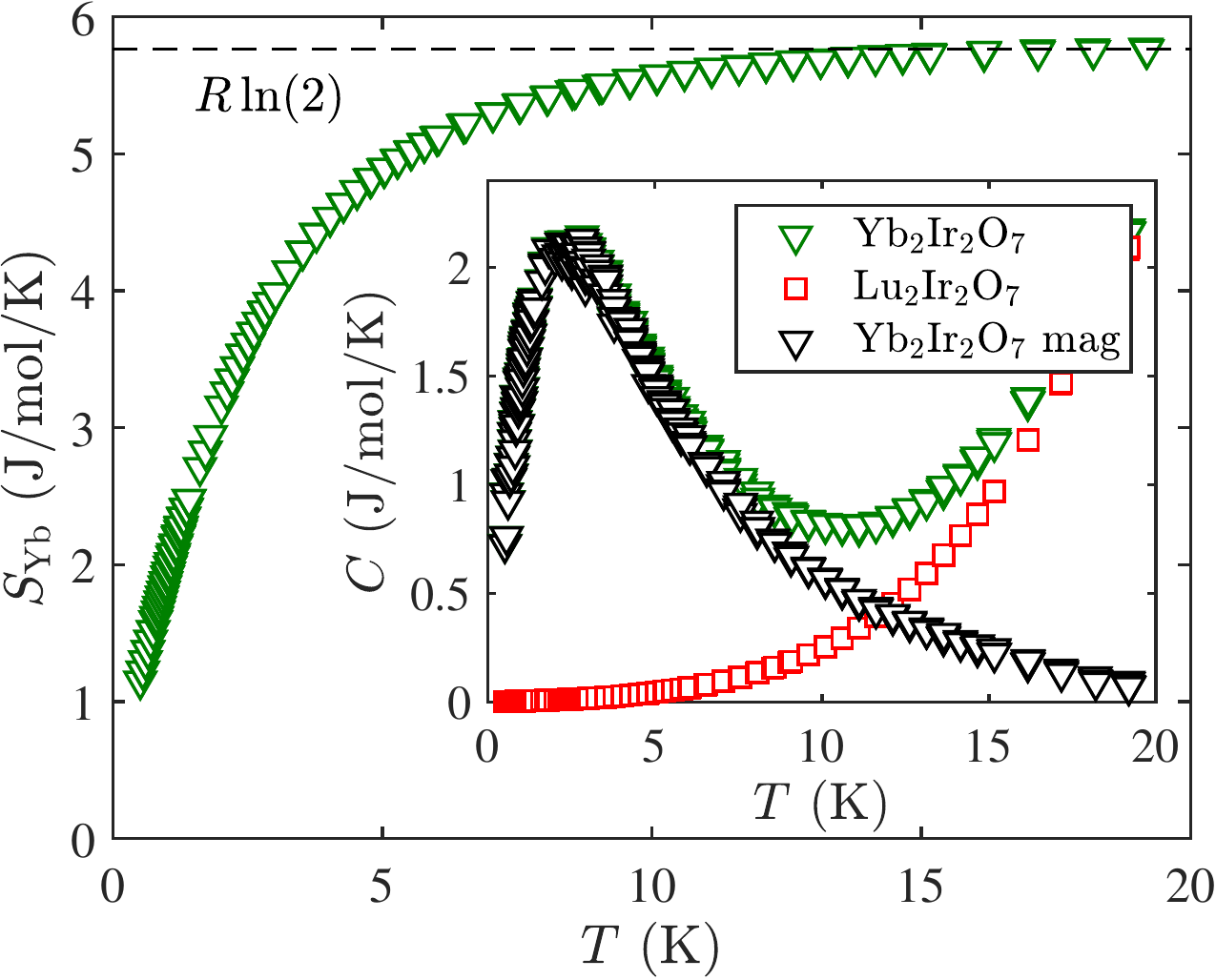}
\caption{Heat capacity and entropy per mole of Yb for our \Ybtts{} sample. }\label{fig:Fig8}
\end{figure}

In other Yb pyrochlores, a broad anomaly in heat capacity at low temperatures has been found to be a signature of low energy spin excitations \cite{Hallas2017,Dun2013,Dun2015,Cai2016,Yaouanc2013,Hodges2002}. The heat capacity of our samples down to $T= 0.5$\,K is shown in the inset to figure \ref{fig:Fig8}. The lattice contribution has been subtracted using the \Lutts{} sample as background to isolate the heat capacity of the Yb moments. We observe no sharp anomalies, but there is a broad peak centred near 2\,K, indicating the build up of correlations in agreement with our neutron diffraction data and the aforementioned results on other Yb pyrochlores. This behaviour is consistent with the absence of a thermodynamic phase transition.

The entropy, $S_\text{Yb}$ is shown in the main panel of Fig.~\ref{fig:Fig8}. The entropy has been offset by $S_0=1.1$ J/mol/K to account for the heat capacity for $T<0.5$\,K. 
The relatively small value of $S_0$ indicates that the  entropy change associated with any phase transition below 0.5\,K is very small.

\subsection{Mean field analysis}

We now present a mean field analysis of the magnetic phases in \Ybtts{} which provides insight into the observed magnetic order of Yb and the cause of its reduced moment. The appropriate effective spin Hamiltonian is \cite{Yan2017,Lefrancois2017,Lefrancois2019}
\begin{align}
    \mathcal{H}&=\sum_{\langle i,j\rangle} \sum_{\mu,\nu} J_{ij}^{\mu\nu} S_i^\mu   S_j^\nu+ \sum_{\langle i,m\rangle} J^\text{Ir-Yb} \SSS_i \cdot \SSS_m, \label{eq:Eq2}
    \end{align}   
where $\SSS_{i,j}$ and $\SSS_m$ are effective spins on the Yb and Ir sites, respectively, $J^\text{Ir-Yb}$ is the coupling between Ir and Yb sites (assumed isotropic), and  $J_{ij}^{\mu \nu} (\mu,\nu=x,y,z)$ is the Yb--Yb exchange matrix, which has four symmetry-allowed components $J_1$--$J_4$. The summations in Eq.~(\ref{eq:Eq2}) are restricted to nearest-neighbor spin pairs. We fixed the Ir spins to the AIAO configuration, treating $J^\text{Ir-Yb}$ as an effective field.  Taking the values of $J_1$--$J_4$ to be the same as those found for \YbTitts{} (Ref.~\onlinecite{Thompson2017}), we minimized the mean-field energy to obtain the phase diagram as function of $J^\text{Ir-Yb}$ shown in Fig.~\ref{fig:Fig1}(c). Details of the calculations can be found in the Appendix.  
 
There are five phases of interest as a function of increasing $J^\text{Ir-Yb}$, illustrated in Fig.~\ref{fig:Fig1}(a). The splayed ferromagnet (Splayed FM) is the ground state of \YbTitts{} (with a splay angle of 17$^\circ$). The  ``3-in-1-out'' (3I1O) arrangement has one of the spins pointing along the local [111] direction anti-aligned with the molecular field from Ir. The AIAO+XY state has two spins approximately following the AIAO structure, while the other two are XY like, pointing perpendicular to the local [111] axis. The Canted AIAO structure has all spins pointing either in or out of the tetrahedra, but at an angle to the local [111] direction. Finally, as 
$J^\text{Ir-Yb}$ increases, the ground state is the pure AIAO structure induced from the Ir sublattice.  Other states are possible depending on the exact values of $J$ and $J^\text{Ir-Yb}$. 

Each state can be decomposed into a sum of the order parameters $\mmm_{k}$ for the five irreps. Fig.~\ref{fig:Fig1}(c) shows $\mmm^2_{k}$ for each of the five irreps as a function of $J^\text{Ir-Yb}$. 

Our observation of a large ferromagnetic component, with up to $\sim10$\% AIAO component and no other components, is consistent with the Splayed FM and 3I1O phases. The lack of any detectable $\Gamma_6$ component in our neutron diffraction data points to the 3I1O phase, except that this phase requires a $\Gamma_{10,B}$ component that is larger than the upper limit placed on it from our neutron data. This is not a significant concern, however, because the amount of $\Gamma_{10,B}$ component is mainly determined by the magnitude of $J_3$, and $J_3$ could well be smaller in \Ybtts{} than in \YbTitts{}. 

Other pyrochlore iridates fit well within this general phase diagram, although details of their ground states vary.
The behavior of the pyrochlore iridates is governed by the relative strength of three interactions: The interaction between (i) the Ir sites (Ir--Ir), (ii) the Ir and the $A$ site (Ir--$A$) and (iii) the $A$ sites ($A$--$A$). The Ir sublattice develops AIAO order at temperatures higher than or equal to the ordering temperature for the $A$ site, indicating that the Ir--Ir interaction is strongest. The Ir--$A$ interaction is typically the second strongest, as evidenced by the onset of induced AIAO order on the $A$ sublattice either simultaneously with the Ir order (\Ndtts{} \cite{Guo2016}, \Tbtts{} \cite{Guo2017}) or at slightly lower temperatures (\Hotts{} \cite{Lefrancois2017}, \Gdtts{} \cite{Lefrancois2019}). The $A$--$A$ interaction is the weakest and only relevant at low temperatures as observed e.g.~in \Tbtts{} where the moments cant slightly below 10\,K \cite{Guo2017}.

\section{Discussion}

In \Ybtts{}, the Ir spins order at 150\,K, while magnetic order of the Yb moments sets in at significantly lower temperatures, between 20\,K and 1.5\,K. We do not have data between these temperatures, but a $\mu$SR study  found a change in behavior at $T^*=20$\,K, indicating the onset of Yb magnetic order\cite{Disseler2012a}.  This indicates that the Ir--Yb and Yb--Yb interactions are the same order of magnitude in our sample, leading to competition.

We have found that the ordered moment on the Yb sites in \Ybtts{} is
$0.57(2)$\,$\mu_{\rm B}$, which is only about 60\% of the $\simeq$0.9$\mu_{\rm B}$ moment found in \YbTitts{} \cite{Yaouanc2016,Hodges2002,Yasui2003}, and only 30\% of the saturated moment $M_s \simeq 1.9$\,$\mu_{\rm B}$ of the ground state Kramers doublet.
To investigate this reduction we have calculated the ordered moment in the mean-field model as function of $J^\text{Ir-Yb}$ by linear spin-wave theory, and give the results in Fig.~\ref{fig:Fig1}(b) as $\langle S \rangle/S$. The ordered moment is found to be suppressed by up to $40\%$ in this model, with the greatest reductions on the phase boundaries. The small ordered moment in our sample implies that it is close to a phase boundary. Even so, the calculated moment reduction is not as large as we observe in \Ybtts{}, and as the mean-field model is semi-classical this suggests that quantum fluctuations further destabilize the order. 

Furthermore, although neutron diffraction detects magnetic order of Yb up to at least $1.5$\,K, AC susceptibility shows no order above $\sim 0.4$\,K (Fig.~\ref{fig:Fig7}). These disparate results from the two techniques can be explained by their different time scales. The time scale for AC susceptibility measurements ($\sim$ $10^{-1}$\,s) is much longer than that of neutron scattering ($\sim$ $10^{-12}$\,s). The neutron thus sees moments which fluctuate more slowly than $\sim$ $10^{-12}$\,s as static. At temperatures between $\sim 0.4$ and  1.5\,K, therefore, the magnetic moments appear static and ferromagnetically correlated over relatively long distances and times to neutrons, but appear dynamic to AC susceptibility. The lack of magnetic hysteresis even at the lowest temperatures is a further indication for the presence of significant spin fluctuations at the lowest temperatures.

The Yb$^{3+}$ ions in \Ybtts{} and \YbTitts{} have weak planar single-ion anisotropy from the crystal field. From our analysis of the neutron spectrum of \Ybtts{} we found that the  the $g$ tensor components parallel and perpendicular to the local $<$111$>$ axes are $g_\parallel = 2.3$ and $g_\perp = 4.0$. The anisotropy ratio $g_\parallel/g_\perp  = 1.7$ for \Ybtts{} is slightly smaller than that for \YbTitts{},  $g_\parallel/g_\perp  = 2.0$. This anisotropy competes with the exchange interactions between the Yb$^{3+}$ ions in \YbTitts{} which favor a splayed ferromagnetic phase that is very close to several other phase boundaries. The resulting phase competition, which is also evidenced by the anomalously reduced ordered moment, is arguably the key to understanding the properties of this compound \cite{Yan2017}.

By replacing the nonmagnetic Ti$^{4+}$ ions in \YbTitts{} with Ir$^{4+}$  having AIAO magnetic order in \Ybtts{} we introduce a weak effective field along the local $<$111$>$ directions on each Yb site which is in direct competition to the other magnetic interactions already present. The effect of this molecular field, as we have established, is to destabilize order and increase frustration. 

\section{conclusions}

In summary, we have found that the Ir sites in \Ybtts{} and \Lutts{} order in the AIAO magnetic structure below $T_{\rm N} \simeq 150$\,K with an ordered moment of around 0.45\,$\mu_{\rm B}$. Upon cooling below $\sim1.5$\,K, the Yb moments in \Ybtts{} begin to order with a dominant ferromagnetic component.
The ordered moment on Yb at the lowest accessible temperature of 40\,mK is only about 30\% of that expected for the ground state Kramers doublet of Yb$^{3+}$, and so the majority of the Yb moment remains dynamic. Our analysis suggests that the suppression of Yb magnetic order in \Ybtts{} is the result of competition between different ground states favored by Yb--Yb exchange, single-ion anisotropy, and the staggered field from the Ir magnetic order. This study demonstrates that the introduction of magnetic ions on the $B$ sites in $A_2B_2$O$_7$ can provide a route to unconventional quantum ground states on the $A$ sites. 

\begin{acknowledgments}
This work was supported by the U.K. Engineering and Physical Sciences Research Council (Grant nos. EP/N034872/1 and EP/N034694/1). Experiments at the ISIS Neutron and Muon Source were supported by a beam-time allocation from the Science and Technology Facilities Council. We thank Russell Ewings for help with the MAPS experiment, and Nic Shannon for informative discussions.
\end{acknowledgments}


%

\clearpage

\section{Sample characterisation}
Here we  present resistivity measurements of our polycrystalline samples of \Lutts{} and \Ybtts{}.  

\begin{figure}
\centering
\includegraphics[width=0.48\textwidth]{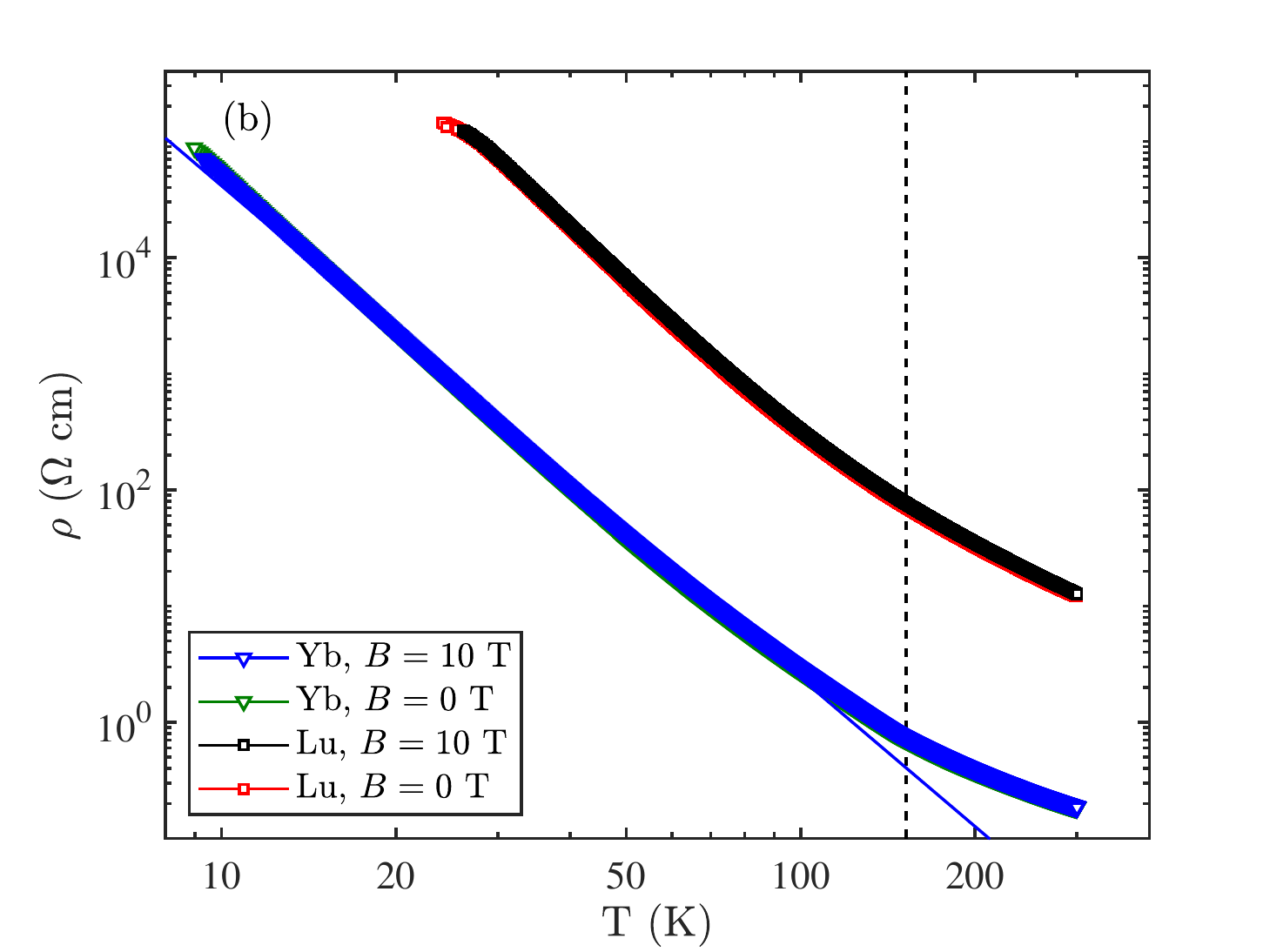}
\caption{Resistivity of \Lutts{} and \Ybtts{} in zero field and in an applied field of 10\,T. The dashed line indicates the magnetic ordering temperature from the magnetization measurements. Below $\sim$25\,K the resistivity of the \Lutts{} sample is higher than the instrument limit.
}
\label{fig:Fig9}
\end{figure}

In Fig.~\ref{fig:Fig9}, the resistivity is seen to increase monotonically upon cooling, from 12\,$\Omega$\:cm (0.2\,$\Omega$\:cm) at 300\,K to more than $10^5$\,$\Omega$\:cm at 25\,K (10\,K) K for the Lu (Yb) sample, in agreement with previous results \cite{Disseler2012a}. In both samples, a gradual crossover from $\rho \sim T^{-2}$ to $\rho \sim T^{-4}$ takes place upon cooling below $\sim$150\,K, although no clear transition is apparent in neither the resistivity nor its derivative. The resistivity does not change significantly on application of a magnetic field of 10\,T.

\section{Normalization of neutron diffraction data in dilution fridge}

Isolation of the Yb magnetic diffraction signal from the structural and Ir magnetic signals requires care. Ideally, we should like to subtract a dataset measured in the paramagnetic phase from data at the lowest temperature to remove the nuclear scattering signal. Unfortunately, the highest temperature accessible in the dilution fridge was 10\,K, which is much lower than the ordering temperature of the Ir sublattice (160\,K). Moreover, the different amounts of attenuation from the Al and Cu sample containers used in the two experiments rules out a simple subtraction of the high temperature data measured in the orange cryostat from the dilution fridge data. 

In order to have confidence in the results, we have used two different methods to isolate the Yb magnetic diffraction. In the following, $I^{(1)}(T)$ and $I^{(2)}(T)$ represent diffraction data measured at temperature $T$ in the orange cryotat and dilution fridge, respectively.

\begin{figure}
    \centering
    \includegraphics[width=0.48\textwidth]{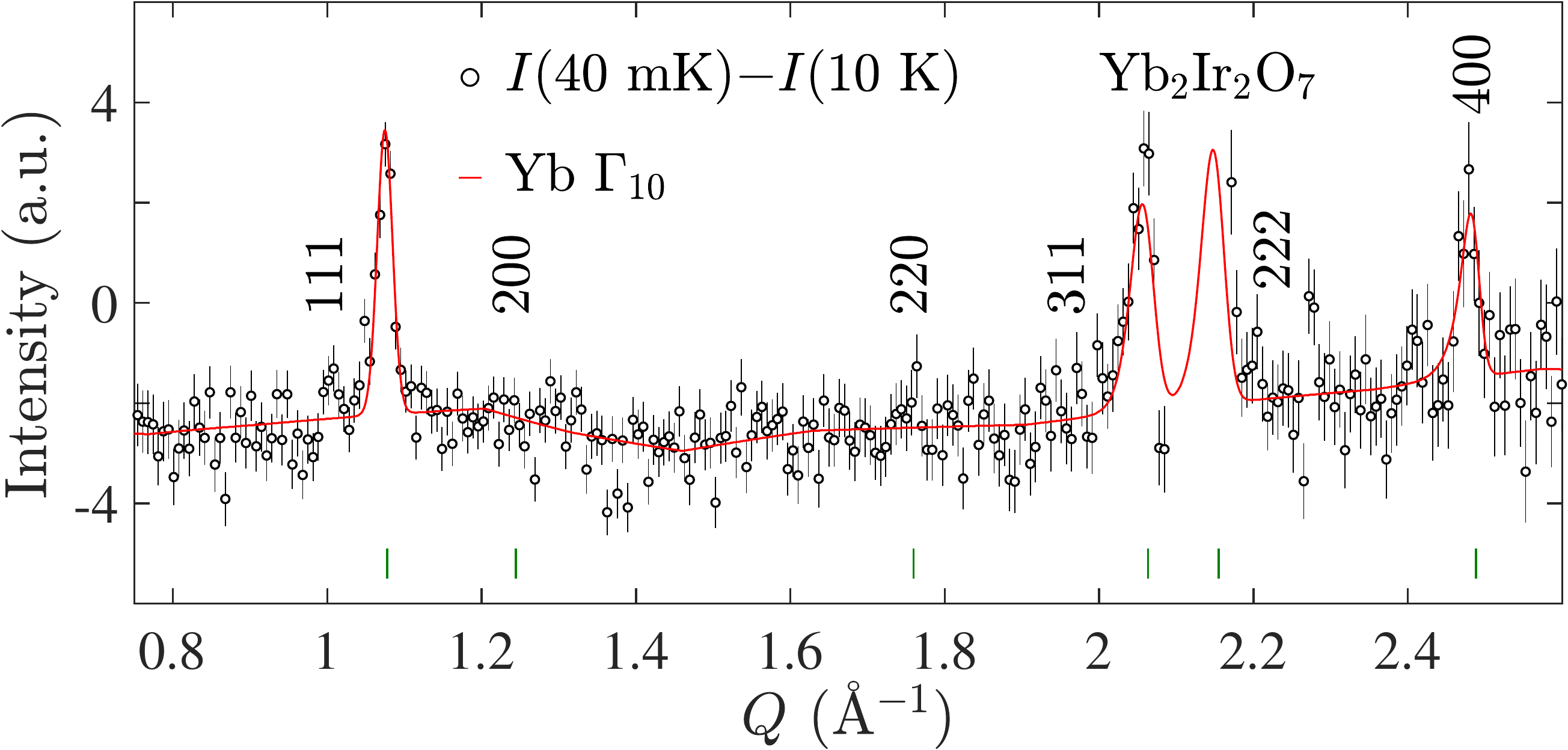}
    \caption{ $I_{\rm mag}^{\text{40 mK}}(\text{Yb})$ calculated using Eq.~(\ref{eq:Eq3}), along with a refinement of the Yb magnetic structure.}
    \label{fig:Fig10}
\end{figure}

The first method is to neglect the Ir sublattice magnetic order and to calculate
\begin{align}
 I_{\rm mag}^{\text{40 mK}}(\text{Yb}) &\approx I^{(2)}(\text{40 mK}) - I^{(2)}(\text{10 K}). \label{eq:Eq3}
\end{align}
This subtraction gives the magnetic signal from the Yb sublattice together with any interference scattering between the Ir and Yb magnetic order. As we show in the main paper, the two sublattices order in different irreps which are independent, and hence the Yb--Ir interference scattering is negligible. In this special case, therefore, the subtraction of the 10\,K data from the 40\,mK data leaves only the magnetic signal from the Yb sublattice, shown in Fig.~\ref{fig:Fig10}. The Yb magnetic structure refinement shown in Fig.~\ref{fig:Fig10} gives the ordered moment to be 0.58(2)~$\mu_\text{B}$/Yb.

\begin{figure}
 \includegraphics[width=0.48\textwidth]{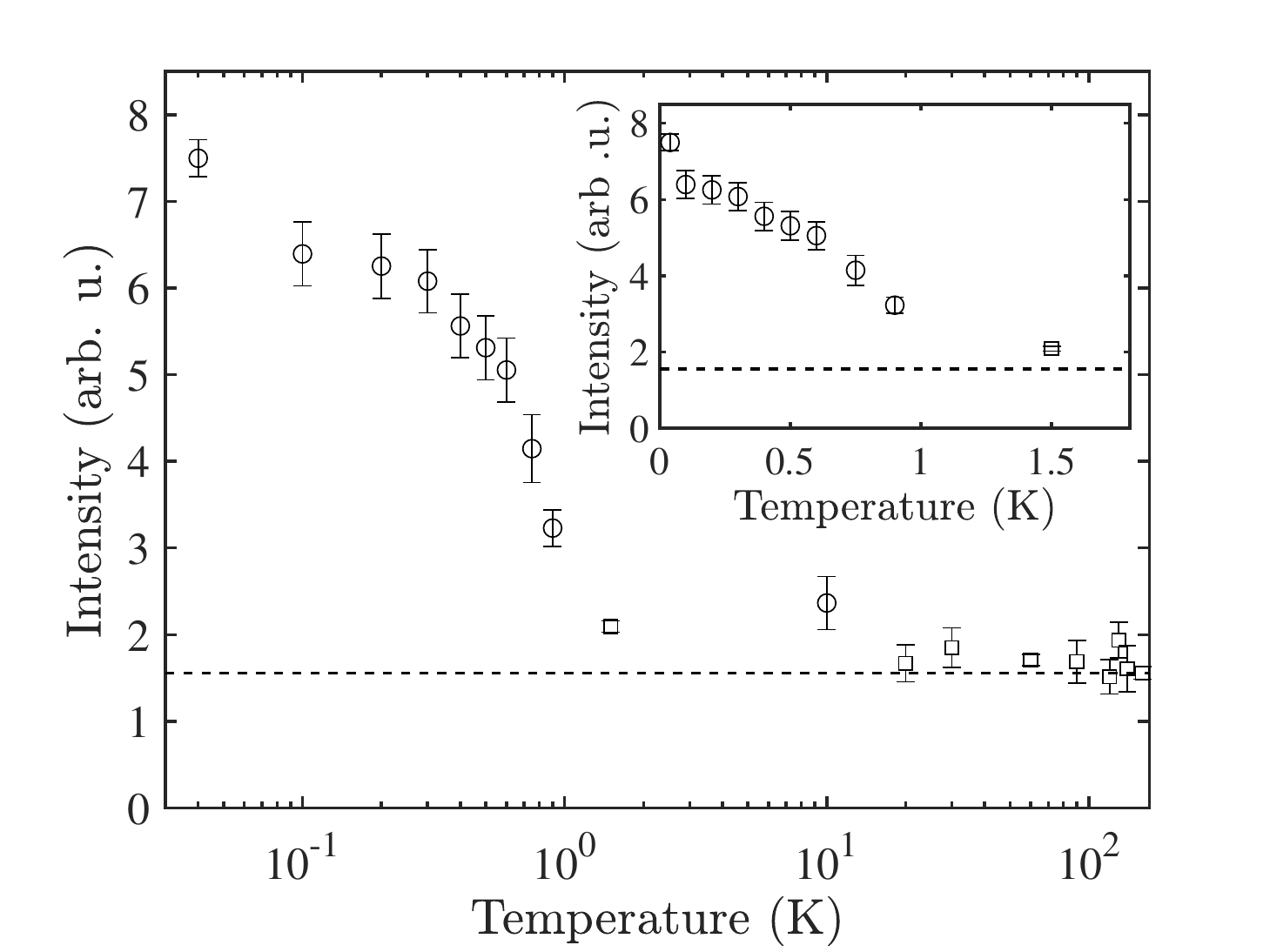}
 \caption{
 The integrated intensity of the (mainly magnetic) 111 reflection as a function of temperature on a log scale. The inset shows the intensity at low temperature on a linear scale. The intensity has been divided by the integrated intensity of the closest strong nuclear peak which is from the 222 reflection. The dashed line shows the intensity of the peak at 160\,K where only nuclear scattering is present. The large error on the 10\,K point is due to the presence of excess helium around the sample which was present only during the measurement at 10\,K in the dilution fridge.  } \label{fig:Fig11}
\end{figure}

In the second method, we isolate the magnetic signal at 40\,mK by calculating 
\begin{align}
 I_{\rm mag}^{\text{40 mK}} &\approx S \times \{I^{(2)}(\text{40 mK}) - I^{(2)}(\text{10 K})\} \nonumber\\
 & + \{I^{(1)}(\text{60 K}) - I^{(1)}(\text{160 K})\},
 \label{eq:Eq4}
\end{align}
where $S = 2.22(4)$ is a scaling factor between the two experiments, calculated from refinements of the nuclear scattering peak intensities. The quoted error is calculated from the uncertainty in the scale factor of the individual refinements at 40 mK and 60 K.
Approximation (\ref{eq:Eq4}) is valid providing the nuclear and magnetic signals satisfy $S \times I^{(2)}(\text{10 K}) \approx I^{(1)}(\text{60 K})$, which applies here because (a) the structural Bragg peaks changes very little between 10 and 60\,K, and (b) the refined Ir magnetic moment is virtually constant over the  range 10 to 60\,K, as may be seen from Fig.~\ref{fig:Fig6} in the main text. To provide an additional check, we examined the integrated intensity of the 111 reflection, which has a weak nuclear component at all temperatures and a magnetic component which develops as the temperature is reduced below $\sim1.5$\,K, see Fig.~\ref{fig:Fig11}. The integrated intensity is seen to remain constant for temperatures between $\sim10$ and 160\,K. The value of the refined magnetic moment obtained using this method is 0.57(2)~$\mu_\text{B}$/Yb, which is consistent with the value obtained from the first method.

\section{Crystal field excitations in \Ybtts{}}

In order to isolate the single-ion magnetic response of Yb we have measured the spectrum of crystal-field excitations within the $^2F_{7/2}$ term of Yb$^{3+}$ ($4f^{13}$) by inelastic neutron scattering. We start with a discussion of previous results in \YbTitts{} as a starting point for our analysis.

Inelastic neutron scattering measurements on \YbTitts{} revealed crystal field excitations at 76, 82 and 116\,meV. \cite{Gaudet2015} The model used in Ref.~\onlinecite{Gaudet2015} uses the Stevens operator formalism and thus includes only the $J=7/2$ manifold of states. Although this is a very good approximation for Yb$^{3+}$, for a proper comparison with our \Ybtts{} data we would like to use the same model for both data sets. 

We therefore fitted the data in Ref.~\onlinecite{Gaudet2015} using the  model \eqref{eq:Eq5}. With only five observables (3 peak positions, 2 intensity ratios) we must fix one of the six crystal field parameters. We choose to fix the highest order parameter $B_6^6$ to the value found in Ref.~\onlinecite{Gaudet2015}. We find that with minor adjustments in the remaining parameters we can obtain a model that fits the data equally as well as the simpler model used in Ref.~\onlinecite{Gaudet2015}. Note that to achieve good agreement with the observed spectrum we needed to include a phonon peak centred near 76\,meV. The best-fit parameters are given in Table~\ref{tab:Tab1}.

\begin{table*}[]
    \centering
\begin{tabular}{c| c c c c c c c c}
\hline\hline	 &$B_2^0$ & $B_4^0$ & $B_4^3$ & $B_6^0$ & $B_6^3$ & $B_6^6$  & $g_\perp$ & $g_\parallel$ \\  \hline 
	\YbTitts{} (Ref.~\onlinecite{Gaudet2015})  &  71.5 & 284.0 &  61.5 & 118.9 & -195.2 &  35.6 &  3.69&  1.92 \\
	\YbTitts{} (This work)  &  71.8 & 284.0 &  47.9 & 119.0 & -195.0 &  35.6   &3.87&  1.94 \\
	Scaled from Ref.~\onlinecite{Princep2015} & 52.7 & 292.8 & 101.3 & 78.5 & -78.5 & 82.6 & 3.97  & 2.40 \\
	\Ybtts{} (This work)  &  72.6 & 258.0 & 116.0 &  84.1 & -99.8 &  82.6  &  4.03&  2.32\\
\hline\hline\end{tabular}
    \caption{Crystal field parameters and components of the $g$ tensor in different models for Yb$_2$Ti$_2$O$_7$ and \Ybtts{}, as described in the text.}
    \label{tab:Tab1}
\end{table*}

We now turn to our experiments. A standard vanadium sample was measured to  normalize the data from runs with different $E_{\rm i}$ and to calibrate the spectra in units of mb\,sr$^{-1}$meV$^{-1}$Yb$^{-1}$. However, an accurate absolute calibration proved not to be possible because of the large neutron absorption cross-section of Ir. With a sample mass of 5.0\,g the average thickness of the sample in the aluminium sachet is $t\simeq 0.08$\,mm giving a nominal $\sim5\%$ absorption according to Beer's law. In the analysis detailed below, however, we find all the calculated intensities to be about a factor of 2 times larger than the observed intensities, indicating much stronger absorption than in the ideal case. This is most likely due to the difficulty in spreading the powder evenly in such a thin layer. 

The normalized spectra were corrected for sample absorption assuming an evenly loaded sample, and for the magnetic form factor of Yb$^{3+}$, $f^2(Q)$, as well as for a small offset on the energy axis.

Figure~\ref{fig:Fig3}(a) is a color map of the corrected intensity as function of scattering vector, $Q$, and energy transfer, $E$. We made a constant-$Q$ cut through the data, averaging the intensity over $3.5<Q<4.5$\,\AA{}$^{-1}$. These cuts are shown in Figs.~\ref{fig:Fig3}(b) and (c) for $E_i=200$ and $110$\,meV, respectively. 

We identify two clear peaks from crystal field excitations at 76.6(6) and 113.5(3)\,meV. The peak at 76.6\,meV has shoulders on both sides which can be modelled with peaks centred near 71 and 81\,meV. We expect one of these to be a phonon and the other a crystal field excitation. We are able to find satisfactory fits to our data for both of these cases. The best fit is found when we attribute the peak near 81\,meV to  the crystal field excitation, although we cannot rule out the other possibility with certainty. For this fit, as well as others described below, we used an approximation of the resolution function to describe the crystal field excitations, and Gaussian functions for the phonons.

We find good agreement between the data for Yb$_2$Ir$_2$O$_7$ and the model with fixed $B_6^6=35.6$\,meV, the value found for Yb$_2$Ti$_2$O$_7$ in Ref.~\onlinecite{Gaudet2015}. However, this model underestimates the susceptibility and the saturated magnetization of Yb$_2$Ir$_2$O$_7$. We then performed fits with other values of $B_6^6$ and found that good fits could be found for a range of values of $B_6^6$. Larger values of $B_6^6$ were found to give better agreement with the susceptibility. Indeed, a scaling of the crystal field parameters from other heavier lanthanides  (Tb$^{3+}$ in Tb$_2$Ti$_2$O$_7$, Ref.~\onlinecite{Princep2015}, Ho$^{3+}$ in Ho$_2$Ti$_2$O$_7$, Ref.~\onlinecite{Rosenkranz2000}, and Er$^{3+}$ in Er$_2$Ti$_2$O$_7$, Ref.~\onlinecite{Gaudet2018a}) to Yb$^{3+}$ gives $B_6^6$ values significantly larger than 36.5\,meV.

We therefore repeated our analysis, fixing $B_6^6=82.6$\,meV, which is obtained by scaling from Tb$^{3+}$ to Yb$^{3+}$ using their respective $4f$ radial averages plus an additional factor of 1.2 to match the overall crystal splitting in Yb$_2$Ti$_2$O$_7$. The fit using this procedure is shown in Figs.~\ref{fig:Fig3}(b)--(c).  We find very good agreement with the data. The susceptibility and magnetisation calculated with this model are shown in Fig.~\ref{fig:Fig2}. The model still slightly underestimates the measurements, but is significantly better than the model with $B_6^6=36.5$\,meV. The remaining discrepancy could be an effect of the exchange interactions which our single-ion model does not take into account.

In total, we have fitted our data using four different models: we have used $B_6^6=35.6$\,meV and $B_6^6=82.6$\,meV, and in each case we have performed the fit assuming, first, the 70\,meV peak and second, the 81\,meV peak is the crystal field excitation. In all models we find $g_\parallel\approx 2$ and $g_\perp \approx 4$. The best fit is found using the method described in detail above. The parameters for this fit are listed in  Table.~\ref{tab:Tab1}.
In this model we find that the components of the wave function from the $J=7/2$ level are a doublet consisting of
\begin{widetext}
\begin{align}
    \psi&=0.904|\tfrac{7}{2},\pm\tfrac{1}{2}\rangle\pm0.413|\tfrac{7}{2},\pm\tfrac{7}{2}\rangle\mp0.094|\tfrac{7}{2},\mp\tfrac{5}{2}\rangle\pm0.037|\tfrac{5}{2},\pm\tfrac{1}{2}\rangle-0.03|\tfrac{5}{2},\mp\tfrac{5}{2}\rangle\pm0.005|\tfrac{7}{2},\mp\tfrac{1}{2}\rangle.
\end{align}
\end{widetext}

\clearpage

\section{Phase diagram}
We here expand on the calculations of the phase diagram. The starting point is the Hamiltonian, Eq.~\eqref{eq:Eq2}. As the magnetic structure has propagation vector $\kkk=0$, the Hamiltonian for the Yb sites reduces to a sum over individual tetrahedra \cite{Yan2017}. The Ir--Yb exchange interaction can be modelled as an effective field which for Yb site $i$ is given by
\begin{align}
    \BBB^\text{Ir-Yb}_i=\sum_i J^\text{Ir-Yb} \SSS_i^\text{Ir}=  2J^\text{Ir-Yb}S^\text{Ir} \hat{\zzz}_i,
\end{align}
where $\hat{\zzz}_i$ is a unit vector along the local (111) direction. We have assumed that the Ir sublattice orders in the AIAO magnetic structure, and that the Ir--Yb exchange interaction is isotropic. 

The Hamiltonian for a single tetrahedron is \cite{Yan2017,Lefrancois2017,Lefrancois2019}
\begin{align}
    \mathcal{H}&=\sum_{\langle i,j\rangle} \sum_{\mu,\nu} \SSS_i^\mu J_{ij}^{\mu\nu}  \SSS_j^\nu+ \sum_{\langle i,m\rangle} J^\text{Ir-Yb} \SSS_i \cdot \SSS_m^\text{Ir}\\
    &=    \sum_{k} \frac{1}{2}a_{k} \mmm_{k}^2 + \frac{1}{2}a_{10,AB} \mmm_{10,A}\cdot \mmm_{10,B} \nonumber\\
    &+ B^\text{Ir-Yb} m_{3}, \label{eq:Eq5}
\end{align}

 where the matrix $\mathbf{J}$ for the 6 bonds in a tetrahedron can be written as 
\begin{align}
    \begin{array}{l}{\mathbf{J}_{01}=\left(\begin{array}{ccc}{J_{2}} & {J_{4}} & {J_{4}} \\ {-J_{4}} & {J_{1}} & {J_{3}} \\ {-J_{4}} & {J_{3}} & {J_{1}}\end{array}\right), \quad \mathbf{J}_{02}=\left(\begin{array}{ccc}{J_{1}} & {-J_{4}} & {J_{3}} \\ {J_{4}} & {J_{2}} & {J_{4}} \\ {J_{3}} & {-J_{4}} & {J_{1}}\end{array}\right)} \\ {\mathbf{J}_{03}=\left(\begin{array}{ccc}{J_{1}} & {J_{3}} & {-J_{4}} \\ {J_{3}} & {J_{1}} & {-J_{4}} \\ {J_{4}} & {J_{4}} & {J_{2}}\end{array}\right), \quad \mathbf{J}_{12}=\left(\begin{array}{ccc}{J_{1}} & {-J_{1}} & {-J_{4}} \\ {-J_{3}} & {J_{1}} & {-J_{4}} \\ {-J_{4}} & {J_{4}} & {J_{2}}\end{array}\right)}\end{array} \nonumber\\
    \begin{array}{l}{\mathbf{J}_{13}=\left(\begin{array}{ccc}{J_{1}} & {J_{4}} & {-J_{3}} \\ {-J_{4}} & {J_{2}} & {J_{4}} \\ {-J_{3}} & {-J_{4}} & {J_{1}}\end{array}\right)}, \quad {\mathbf{J}_{23}=\left(\begin{array}{ccc}{J_{2}} & {-J_{4}} & {J_{4}} \\ {J_{4}} & {J_{1}} & {-J_{3}} \\ {-J_{4}} & {-J_{3}} & {J_{1}}\end{array}\right)}\end{array}.
\end{align}
Site and bond labels follow the conventions in Refs.~\onlinecite{Yan2017,Ross2011}. The coefficients $a_{k}$ are given by
\begin{align} 
a_{3} &=-2 J_{1}+J_{2}-2\left(J_{3}+2 J_{4}\right), \\ 
 a_{6} &=-2 J_{1}+J_{2}+J_{3}+2 J_{4}, \\
 a_{8} &=-J_{2}+J_{3}-2 J_{4}, \\ 
 a_{10,A} &=2 J_{1}+J_{2}, \\
 a_{10,B} &=-J_{2}-J_{3}+2 J_{4}, \\
 a_{10,AB} &=-\sqrt{8} J_{3},
\end{align}
 and 
$\mmm_{k}$ are order parameters associated with the five different types of ordered phases \cite{Yan2017}, as also described in the main text. Their precise definitions can be found in Table~\ref{tab:Tab2}, which is a copy of Table III of Ref.~\onlinecite{Yan2017}. In Ref.~\onlinecite{Yan2017} the states are named after which point groups they transform under; here we name them using the irrep naming convention used in Refs.~\onlinecite{Guo2016,Guo2017}. 
We note that we have merely rewritten the Hamiltonian in a different form; no approximations have been made at this point. 
\begin{table*}[]
    \centering
    \begin{tabular}{ccc}
    \hline
    \hline
    Order parameter & Definition in terms of spin components & Associated ordered phases \\ \hline
$m_{\Gamma_3}$& $\frac{1}{2\sqrt{3}}\left(S_{0}^{x}+S_{0}^{y}+S_{0}^{z}+S_{1}^{x}-S_{1}^{y}-S_{1}^{z}-S_{2}^{x}+S_{2}^{y}-S_{2}^{z}-S_{3}^{x}-S_{3}^{z}+S_{3}^{z}\right)$ & All-in-all-out \\
$\mmm_{\Gamma_6}$ & $\left(\begin{array}{c}{\frac{1}{2 \sqrt{6}}\left(-2 S_{0}^{x}+S_{0}^{y}+S_{0}^{z}-2 S_{1}^{x}-S_{1}^{y}-S_{1}^{z}+S S_{2}^{x}+S_{2}^{y}-S_{2}^{z}+2 S_{3}^{x}-S_{3}^{z}+S_{3}^{z}\right)} \\ {\frac{1}{2 \sqrt{2}}\left(-S_{0}^{y}+S_{0}^{z}+S_{1}^{y}-S_{1}^{z}-S_{2}^{y}-S_{2}^{z}+S_{3}^{y}+S_{3}^{z}\right)}\end{array}\right)$ & $\psi_2$ and $\psi_3$ \\
$\mmm_{\Gamma_8}$  & $\left(\begin{array}{c}{\frac{1}{2 \sqrt{2}}\left(-S_{0}^{y}+S_{0}^{z}+S_{1}^{y}-S_{1}^{z}+S_{2}^{y}+S_{2}^{z}-S_{3}^{y}-S_{3}^{z}\right)} \\ {\frac{1}{2 \sqrt{2}}\left(S_{0}^{x}-S_{0}^{z}-S_{1}^{x}-S_{1}^{z}-S_{2}^{x}+S_{2}^{z}+S_{3}^{x}+S_{3}^{z}\right)} \\ {\frac{1}{2 \sqrt{2}}\left(-S_{0}^{x}+S_{0}^{y}+S_{1}^{x}+S_{1}^{y}-S_{2}^{x}-S_{2}^{y}+S_{3}^{x}-S_{3}^{y}\right)}\end{array}\right)$ & Palmer-Chalker \\
$\mmm_{\Gamma_{10,A}}$  & $\left(\begin{array}{c}{\frac{1}{2}\left(S_{0}^{x}+S_{1}^{x}+S_{2}^{x}+S_{3}^{x}\right)} \\ {\frac{1}{2}\left(S_{0}^{y}+S_{1}^{y}+S_{2}^{y}+S_{3}^{y}\right)} \\ {\frac{1}{2}\left(S_{0}^{z}+S_{1}^{z}+S_{2}^{z}+S_{3}^{z}\right)}\end{array}\right)$ & Ferromagnet \\
$\mmm_{\Gamma_{10,B}}$  & $\left(\begin{array}{l}{\frac{-1}{2\sqrt{2}}\left(S_{0}^{y}+S_{0}^{z}-S_{1}^{y}-S_{1}^{z}-S_{2}^{y}+S_{2}^{z}+S_{3}^{y}-S_{3}^{z}\right)} \\ {\frac{-1}{2 \sqrt{2}}\left(S_{0}^{x}+S_{0}^{z}-S_{1}^{x}+S_{1}^{z}-S_{2}^{x}-S_{2}^{z}+S_{3}^{x}-S_{3}^{z}\right)} \\ {\frac{-1}{2 \sqrt{2}}\left(S_{0}^{x}+S_{0}^{y}-S_{1}^{x}+S_{1}^{y}+S_{2}^{x}-S_{2}^{y}-S_{3}^{x}-S_{3}^{y}\right)}\end{array}\right)$ & Noncollinear ferromagnet \\
\hline \hline
    \end{tabular}
    \caption{Definitions of symbols used in Eq.~\eqref{eq:Eq5}. Each $\mmm_{\Gamma_k}$ is the order parameter associated with the phase given in the right column. The table is reproduced from Ref.~\onlinecite{Yan2017}.}
    \label{tab:Tab2}
\end{table*}

When $B^\text{Ir-Yb}=0$,  the energy is minimized when one $\mmm_\lambda=1$ and the rest are 0 (except at phase boundaries). This automatically satisfies the physical constraint that the spins must have the same magnitude \cite{Yan2017}.  When $B^\text{Ir-Yb}\neq 0$, the different order parameters mix, and it is not trivial to analytically satisfy the constraint that the spins are normalized. We have therefore investigated the phase diagram of \Ybtts{} as function of $J_\textrm{Ir-Yb}$ numerically, keeping the exchange constants between Yb sites fixed to the values found in \YbTitts{}, \cite{Thompson2017}. We implemented the calculations independently in MATLAB and SpinW\cite{Toth2015} with identical results.
 The resulting phase diagram as a function of $J^\text{Ir-Yb}$ is shown in Fig.~\ref{fig:Fig1} in the main paper.

\end{document}